\documentclass[12pt]{iopart}
\usepackage{times}
\usepackage{cite}
\usepackage{graphicx}

\newcommand{\lwig}{\mbox{\,\raisebox{.3ex}
    {$<$}$\!\!\!\!\!$\raisebox{-.9ex}{$\sim$}\,}}
\newcommand{\gwig}{\mbox{\,\raisebox{.3ex}
    {$>$}$\!\!\!\!\!$\raisebox{-.9ex}{$\sim$}}\,}

\newcommand{\massnophhalo}{$m_\nu=3.46^{+1.73(4.03)}_{-1.34(2.32)}$~eV }
\newcommand{\massnopheg}{$m_\nu=0.20^{+0.19(0.61)}_{-0.12(0.18)}$~eV, }
\newcommand{\massnophno}{$m_\nu=0.40^{+0.32(0.87)}_{-0.16(0.27)}$~eV, }
\newcommand{\lowbdnopheg}{$m_\nu > 0.02$~eV } 
\newcommand{\chiminnophhalo}{$\chi^2_{\rm min}=15.64$ } 
\newcommand{\chiminnopheg}{$\chi^2_{\rm min}=25.82$ } 
\newcommand{\chiminnophno}{$\chi^2_{\rm min}=8.03$, }

\newcommand{\massleehalo}{$m_\nu=3.71^{+1.40(3.27)}_{-1.12(1.96)}$~eV, }
\newcommand{\massleeeg}{$m_\nu =0.77^{+0.48(1.36)}_{-0.30(0.51)}$~eV, }

\newcommand{\massrangehaloonesig}{$2.1$~eV~$\leq~m_\nu\leq~6.7$~eV }
 
\newcommand{\massrangeegonesig}{$0.08$~eV~$\leq~m_\nu\leq~1.3$~eV} 
\newcommand{\massrangenoonesig}{$0.24$~eV~$\leq~m_\nu\leq~2.6$~eV} 

\newcommand{\massrangeegstrongonesig}{$0.08$~eV~$\leq~m_\nu\leq~0.40$~eV} 
\newcommand{\massnophegbestfit}{$m_\nu=0.20$~eV}

\begin{document}

\title{Z-Burst Scenario for the Highest Energy Cosmic Rays*}
\author{Z. Fodor}
\address{Institute for Theoretical Physics, E\"otv\"os University, 
Budapest, P\'azm\'any 1/a, H-1117} 
\author{\footnote[0]{*Talk given
at Beyond the Desert '02, Oulu, Finland, June 2-8, 2002.}S.D. Katz$^\dagger$\footnote[0]{
$^\dagger$On leave from Institute for Theoretical Physics, E\"otv\"os University, 
P\'azm\'any 1, H-1117 Budapest, 
Hungary},  A. Ringwald}
\address{Deutsches Elektronen-Synchrotron DESY, Hamburg, Notkestra\ss e 85, 
D-22607}
\begin{abstract}
The origin of highest energy cosmic rays is yet unknown. 
An appealing possibility is the so-called Z-burst scenario, in which
a large fraction of these cosmic rays are decay products of Z bosons
produced in the scattering of ultrahigh energy neutrinos on cosmological 
relic neutrinos. The comparison between the observed and predicted spectra
constrains the mass of the heaviest neutrino. The required neutrino mass
is fairly robust against variations of the presently unknown quantities, such
as the amount of relic neutrino clustering, the universal photon radio 
background and the extragalactic magnetic field. Considering different 
possibilities for the ordinary cosmic rays the required neutrino masses are
determined. In the most plausible case that the ordinary cosmic rays are of 
extragalactic origin and the universal radio background is strong enough to
suppress high energy photons, the required neutrino mass is 
\massrangeegstrongonesig\ .
The required ultrahigh energy neutrino flux should be detected in the near 
future by experiments such as AMANDA, RICE or the Pierre Auger Observatory.
\end{abstract}

\vspace*{-12.2cm}
\mbox{} \hfill {\small ITP-Budapest 590, DESY 02-145}\\
\vspace*{11.3cm}

\section{Introduction}

The existence of a background gas of
free photons and neutrinos is predicted by cosmology. 
The measured cosmic microwave background (CMB)
radiation supports the applicability of standard cosmology
back to photon decoupling which occured approximately one 
hundred thousand years after the big bang. 
The relic neutrinos have decoupled much earlier, when the universe 
had a temperature of one MeV and the age
of just one second. Thus, a measurement of the relic neutrinos would provide a
new window to the early universe.
The predicted average number density of relic neutrinos per light 
($m_{\nu_i}\ll 1$~MeV) species $i$ is $\simeq$ 56~cm$^{-3}$.
This density is comparable with that of photons. However,
since neutrinos interact only weakly, the relic neutrinos have not yet been 
detected.

\begin{figure}[b]
\begin{center}
\includegraphics[width=5.0cm]
{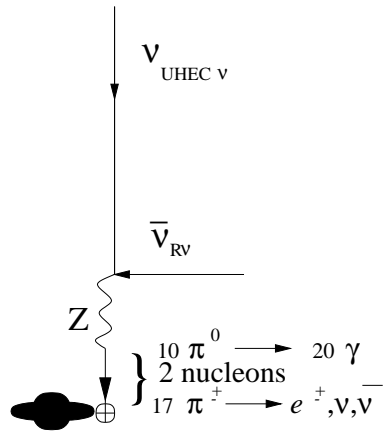}
\caption[...]{\label{illu}
Illustration of a Z-burst resulting from the resonant annihilation of an 
ultrahigh energy cosmic neutrino on a relic (anti-)neutrino. (Adapted from 
Ref. \cite{Pas:2001nd}.)
}
\end{center}
\end{figure}

Recently, an indirect detection possibility for relic 
neutrinos has been discussed~\cite{Weiler:1982qy,%
Roulet:1993pz,Yoshida:1997ie,Fargion:1999ft,Weiler:1999sh}.
It is based on so-called Z-bursts resulting from the 
resonant annihilation of ultrahigh energy cosmic neutrinos
(UHEC$\nu$s) with relic 
neutrinos into $Z$ bo\-sons 
(cf. Fig.~\ref{illu}). The decay products of the $Z$ may be detected and 
yield thereby indirect evidence for the existence of relic neutrinos.

According to recent oscillation 
measurments (see e.g. \cite{Fukuda:1998mi,
Fukuda:2001nj,Ahmad:2001an,Athanassopoulos:1995iw}),  neutrinos have 
non-vanishing
masses, $m_{\nu_i}$. In this case the resonant energy at 
which the $Z$ production 
cross-section is sizeable reads
\begin{eqnarray}
\label{eres}
E_{\nu_i}^{\rm res} = \frac{M_Z^2}{2\,m_{\nu_i}} = 4.2\cdot 10^{21}\ {\rm eV}  
\left( \frac{1\ {\rm eV}}{m_{\nu_i}}\right)
\,,
\end{eqnarray}
where $M_Z$ is the mass of the Z boson. 
For neutrino masses of ${\mathcal O}(1)$~eV the resonant energy is remarkably
close to the energies of the highest energy cosmic rays observed at 
Earth by collaborations such as AGASA~\cite{Takeda:1998ps}, Fly's 
Eye~\cite{Bird:yi,Bird:wp,Bird:1994uy}, 
Haverah Park~\cite{Lawrence:cc,Ave:2000nd}, HiRes~\cite{Kieda00}, and 
Yakutsk~\cite{Efimov91} (for a review, see 
Ref.~\cite{Nagano:2000ve}). 
The possibility that the ultrahigh energy cosmic rays (UHECRs)
may originate from $Z$-bursts was first discussed
in ~\cite{Fargion:1999ft,Weiler:1999sh}.

One of the most outstanding puzzles in cosmic ray physics is the
existence of cosmic ray events above the Greisen-Zatsepin-Kuzmin (GZK) 
cutoff~\cite{Greisen:1966jv,Zatsepin:1966jv}. The energy spectrum of ultrahigh 
energy
cosmic rays is expected to show a cutoff around $4\cdot 10^{19}$~eV which is 
not conclusively seen 
in the data~\cite{Bhattacharjee:2000qc}.
With the help of the Z-burst scenario one can not only detect the relic neutrino 
background indirectly but also explain the existence of the highest energy 
cosmic rays~\cite{Waxman:1998yh,Yoshida:1998it,Blanco-Pillado:2000yb,Gelmini:1999qa,%
Gelmini:2000ds,%
Weiler:1999ny,Crooks:2001jw,Gelmini:2000bn,Pas:2001nd,Fargion:2000pv,%
Fodor:2001qy,Fodor:2001rg,Fodor:2002hy,McKellar:2001hk,Ringwald:2001mx,%
Kalashev:2001sh,Gelmini:2002xy,Pas:2002ff,Singh:2002de}.

The comparison between the observed and predicted UHECR spectra gives
constraints on the mass of the heaviest neutrino~\cite{Fodor:2001qy,%
Fodor:2001rg,Fodor:2002hy,Ringwald:2001mx}.
In this review our quantitative analysis of the Z-burst scenario is presented.
The requried mass of the heaviest neutrino, as well
as the necessary UHE$\nu$ flux is determined. 

The organization of this review is as follows. The next section summarizes
our present knowledge on neutrino masses. Section~\ref{sect:spectra} gives
the produced spectrum of protons and photons from Z-bursts. The comparison 
with the observed spectrum is presented in Section~\ref{sect:determination}
and the conclusion is given in Section~\ref{sect:discussion}.

\section{Neutrino masses} \label{sect:masses}
According to recent experiments, 
neutrinos almost certainly have non-vanishing masses and mixing angles. 
The evidence for neutrino oscillation is compelling 
for atmospheric 
neutrinos~\cite{Fukuda:1998mi}, 
strong for solar 
neutrinos~\cite{Fukuda:2001nj,Ahmad:2001an}, 
and so-far unconfirmed for neutrinos produced in the laboratory and studied by 
e.g. the LSND 
collaboration~\cite{Athanassopoulos:1995iw}.
However, neutrino oscillations are sensitive only to the mass (squared) 
differences 
$\triangle m_{ij}^2 = m_{\nu_i}^2 - m_{\nu_j}^2$, not to the individual masses, 
$(m_{\nu_4}>)\ m_{\nu_3}>m_{\nu_2}>m_{\nu_1}$, themselves.
Only a lower bound on the mass of the heaviest neutrino can be derived from 
these observations, e.g. 
\begin{equation}
\label{lim_low_atm}
m_{\nu_3}\geq\sqrt{\triangle m_{\rm atm}^2}\,\gwig\, 0.04\ {\rm eV}
\end{equation}
from the atmospheric mass splitting in a three neutrino flavour scenario and
\begin{equation}
\label{lim_low_lsnd}
m_{\nu_4}\geq\sqrt{\triangle m_{\rm LSND}^2}\,\gwig\, 0.4\ {\rm eV}
\end{equation}
from the still allowed value of the LSND mass splitting in a four 
flavour scenario.

For an investigation of 
the absolute scale of neutrino masses one has to exploit different types 
of experiments, 
such as the search for mass imprints in the endpoint spectrum of tritium 
beta ($\beta$) decay or the search for 
neutrinoless double beta ($0\nu 2\beta$) decay.  
At present, these direct kinematical measurements of neutrino masses 
provide only upper limits, e.g. 
\begin{equation}
\label{lim_beta}
m_\beta\equiv \sqrt{\sum_i |U_{ei}|^2\, m_{\nu_i}^2 }<2.2\div 2.5\ {\rm eV}
\hspace{2ex} (95\,\%\ {\rm C.L.})
\,,
\end{equation} 
with $U$ being the leptonic mixing matrix,  
for the effective mass measured in tritium $\beta$ 
decay~\cite{Weinheimer:1999tn,Lobashev:1999tp,Bonn:2001tw}, 
and 
\begin{equation}
\langle m_\nu\rangle \equiv \left| \sum_i U_{ei}^2\, m_{\nu_i} \right| 
<0.33\div 1.35\ {\rm eV}
\hspace{2ex} (90\,\%\ {\rm C.L.})
\end{equation}
for the Majorana neutrino mass 
parameter~\cite{Aalseth:1999ji,Klapdor-Kleingrothaus:2001yx}
appearing in $0\nu 2\beta$ 
decay. Recently an evidence for 
neutrinoless double beta decay was 
reported~\cite{Klapdor-Kleingrothaus:2002ke}. 
The corresponding neutrino mass range is:
\begin{equation}
\label{m0n2b}
\langle m_\nu\rangle =(0.11\div 0.56)\ {\rm eV}
\hspace{2ex} (95\,\%\ {\rm C.L.})
\end{equation}
Combining the experimental constraints from oscillations and from tritium 
$\beta$ decay, one infers
upper bounds on the mass of the heaviest neutrino,
\begin{equation}
\label{lim_comb_osc_beta}
m_{\nu_3}<\sqrt{m_\beta^2+\triangle m_{\rm atm}^2}\,\lwig\, 2.5\ {\rm eV}\,, 
\end{equation}
in a three flavour, and 
\begin{equation}
\label{lim_comb_osc_beta_lsnd}
m_{\nu_4}<\sqrt{m_\beta^2+\triangle m_{\rm LSND}^2}\,\lwig\, 3.8\ {\rm eV}\,, 
\end{equation}
in a four flavour scenario. 

Further information on the absolute scale of neutrino masses 
can be obtained through cosmological and astrophysical considerations.
Neutrinos in the $0.1\div 1$~eV mass range have cosmological implications 
since they represent a non-negligible part of dark matter. 
This gives the opportunity to put upper limits
on neutrino masses from 
cosmology~\cite{Gershtein:1966gg,Dolgov:2002wy,Dolgov:2002ab,Hannestad:2002iz}.
A recent galaxy redshift survey gives an upper bound
\begin{equation}
\label{lim_cosm}
\sum_i 
m_{\nu_i} < 1.8\ {\rm eV}
\end{equation}
on the sum of the neutrino 
masses~\cite{Elgaroy:2002bi}.
From the spread of arrival times of neutrinos from supernova SN 1987A, coupled 
with the measured 
neutrino energies,  a time-of-flight limit of  
$m_\beta <23$~eV can be derived~\cite{Loredo:1988mk,Kernan:1995kt}, which, 
however, is not competitive 
with the direct limit~(\ref{lim_beta}). According to a recent study,
leptogenesis requires that all the neutrinos have masses smaller than 
0.2~eV~\cite{Buchmuller:2002jk}.

\section{\label{sect:spectra}Z-burst spectra}

The calculation of the Z-burst spectra of protons and photons
was performed in three steps.
First, we determined the probability
of Z production as a function of the distance from Earth.
Secondly, using results from collider experiments we derived
the energy distribution of the produced protons and photons 
in the laboratory (lab) system.
As a last step the energy losses due to the propagation of the protons and 
photons were taken into account.

The prediction of the differential proton flux, i.e. the number of protons 
arriving at Earth with energy $E$ per units of energy ($E$), of area ($A$), of 
time ($t$) and
of solid angle ($\Omega$),  
\begin{equation}
F_{p|Z} ( E ) = \frac{{\rm d}N_{p|Z}}{{\rm d}E\,{\rm d}A\,{\rm d}t\,{\rm 
d}\Omega}
\,,
\end{equation} 
from Z-bursts  can be summarized as
\begin{eqnarray}
\label{p-flux}
F_{p|Z} ( E ) &= & \sum_{i}
\int\limits_0^\infty {\rm d}E_p \int\limits_0^{R_{\rm max}} {\rm d}r
\\ \nonumber && \times   
\left[  \int\limits_0^\infty {\rm d} E_{\nu_i}\,F_{\nu_i}(E_{\nu_i},r)\, 
n_{\bar\nu_i}(r)
+ \int\limits_0^\infty {\rm d} E_{\bar\nu_i}\, F_{\bar\nu_i}(E_{\bar\nu_i},r)\, 
n_{\nu_i}(r)
\right]
\\[1ex] \nonumber && \times
 \sigma_{\nu_i\bar\nu_i}(s)\,
{\rm Br}(Z\to {\rm hadrons})\,
\frac{{\rm d} N_{p+n}}{{\rm d} E_p}
(-)\frac{\partial}{\partial E} P_p(r,E_p;E)
\,,
\end{eqnarray}
with the following important building blocks:
the UHEC$\nu$ fluxes $F_{\nu_i}(E_{\nu_i},r)$ at the energies 
$E_{\nu_i}$ ($\approx E_{\nu_i}^{\rm res}$) and at the distance $r$
of Z production to Earth,  the number density 
$n_{\nu_i}(r)$ of the 
relic neutrinos, 
the Z production cross section $\sigma_{\nu_i\bar\nu_i}(s)$ at centre-of-mass 
(cm)
energy squared $s = 2\,m_{\nu_i}\,E_{\nu_i}$, the branching ratio 
${\rm Br}(Z\to {\rm hadrons})$, 
the energy distribution ${\rm d} N_{p+n}/{\rm d} E_p$ of the produced protons 
(and neutrons) with energy $E_p$, and the probability $P_p(r,E_p;E)$ 
that a proton created at a distance $r$ with energy $E_p$ arrives
at Earth above the threshold energy $E$. 

A similar expression as Eq.~(\ref{p-flux}) holds for the differential
photon flux from Z-bursts. One simply has to insert the energy distribution
of photons from Z-bursts and the photon propagation function. The latter,
$P_\gamma (r,E_\gamma ;E)$ has a different meaning than that of the proton's.
It gives the
expected number of detected photons above the threshold 
energy $E$ if one photon started from a distance of $r$ with energy $E_\gamma$.

The building blocks related to Z-production 
and decay -- $\sigma_{\nu_i\bar\nu_i}$, 
the hadronic branching ratio, and 
the momentum distributions ${\rm d} N_i/{\rm d} E_i$ -- are very well known. 
The determination of the propagation functions 
$P_i$ -- though CPU intensive -- can be done using the precisely 
known microwave
backround spectrum and the fairly well known radio and infrared 
background spectra. 
The first two ingredients, the flux of UHEC$\nu$s, 
$F_{\nu_i}(E_{\nu_i},r)$, and the radial distribution of
the relic neutrino number density $n_{\nu_i}(r)$ are, however,  much less 
accurately known. 
In the following all these ingredients are discussed in detail.

\subsection{\label{zproddecay}Z production and decay}

The s-channel Z-exchange annihilation cross section into any fermion 
anti-fermion ($f\bar f$) pair 
is given by 
\begin{equation}
\label{z-res-cs-roulet}
\sigma_{\nu_i \bar{\nu}_i}^{Z} (s)
=
\frac{G_F^2\,s}{4\,\pi}\, \frac{M_Z^4}{(s-M_Z^2)^2+M_Z^2\, \Gamma_Z^2} 
\ N_{\rm eff}(s)\,,
\end{equation}
where $s$ is the cm energy squared, $\Gamma_Z$ is the total width of the Z 
boson, 
and $N_{\rm eff}$ is the effective number of annihilation channels,
\begin{eqnarray}
N_{\rm eff}(s) = 
\sum_f \theta (s-4\,m_f^2)
 \, \frac{2}{3}\,n_f \left( 1 - 
8\, T_{3\,f}\, q_f\, \sin^2\theta_W
+8\,q_f^2\sin^4\theta_W\right)
\,.
\end{eqnarray}
Here the sum is over all fermions with $m_f<\sqrt{s}/2$, with charge
$q_f$ (in units of the proton charge), isospin $T_{3\,f}$ ($1/2$ for $u,c$ and 
neutrinos; $-1/2$ for $d,s,b$ and negatively charged leptons), and $n_f=1(3)$ 
for leptons (quarks ($q$)). With $\sin^2\theta_W=0.23147(16)$ for the $\sin^2$ 
of
the effective Weinberg angle and $G_F=1.16639(1)\times 10^{-5}$ GeV$^{-2}$ 
for the Fermi coupling constant~\cite{Groom:2000in}, 
formula~(\ref{z-res-cs-roulet}) gives, at the $Z$-mass,
\begin{eqnarray}
\sigma (\nu_i \bar{\nu}_i \to Z^\ast \to
 {\rm all\ }q\bar q)\mid_{s=M_Z^2} &=&
314.9\ {\rm nb}\,,
\\[1ex] 
\label{cs-tot-zpeak}
\sigma (\nu_i \bar{\nu}_i \to Z^\ast \to
 {\rm all}\ f\bar f)\mid_{s=M_Z^2} &=&
455.6\ {\rm nb}\,,
\end{eqnarray}
with a branching ratio $\sigma (\nu_i \bar{\nu}_i \to Z^\ast
\to {\rm all\ }q\bar q)/\sigma (\nu_i \bar{\nu}_i \to Z^\ast
\to {\rm all}\ f\bar f)\mid_{s=M_Z^2} =  0.6912$, in good agreement
with the experimental result~\cite{Groom:2000in}, 
\begin{equation}
{\rm Br}(Z\to {\rm hadrons})=(69.89\pm 0.07)\,\%\,.
\end{equation}  

Note that the cross-section is sharply peaked at the resonance
cm energy squared $s=M_Z^2$. Correspondingly, it acts like 
a $\delta$-function in 
the integration over the energies $E_{\nu_i}$ in Eqn.~(\ref{p-flux}), 
and we can assume that the UHEC$\nu$ fluxes are 
constant in the relevant energy
region. Thus, introducing the energy-averaged annihilation 
cross section~\cite{Weiler:1982qy,Weiler:1999sh},
\begin{equation}
\label{sig_ann}
\langle \sigma_{\rm ann}\rangle 
\equiv
\int
\frac{{\rm d}s}{M_Z^2}\,\sigma_{\rm ann} (s)
= 2\,\pi\,\sqrt{2}\,G_F
= 40.4\ {\rm nb}\,,  
\end{equation}
which is the effective cross section for all neutrinos within 
$1/2\,\delta E_{\nu_i}^{\rm res}/E_{\nu_i}^{\rm res}=\Gamma_Z/M_Z=2.7\,\%$ of 
their
peak annihilation energy, we can write
\begin{eqnarray}
\label{eresfres}
\int\limits_0^\infty {\rm d} 
E_{\nu_i}\,F_{\nu_i}(E_{\nu_i})\,\sigma_{\nu_i\bar\nu_i}( 
s=2\,m_{\nu_i}\,E_{\nu_i})
\simeq 
E_{\nu_i}^{\rm res}\,F_{\nu_i}(E_{\nu_i}^{\rm res})\,\langle \sigma_{\rm 
ann}\rangle 
\,.
\end{eqnarray}

\begin{figure}
\begin{center}
\includegraphics[bbllx=20pt,bblly=225pt,bburx=570pt,bbury=608pt,width=7.7cm]
{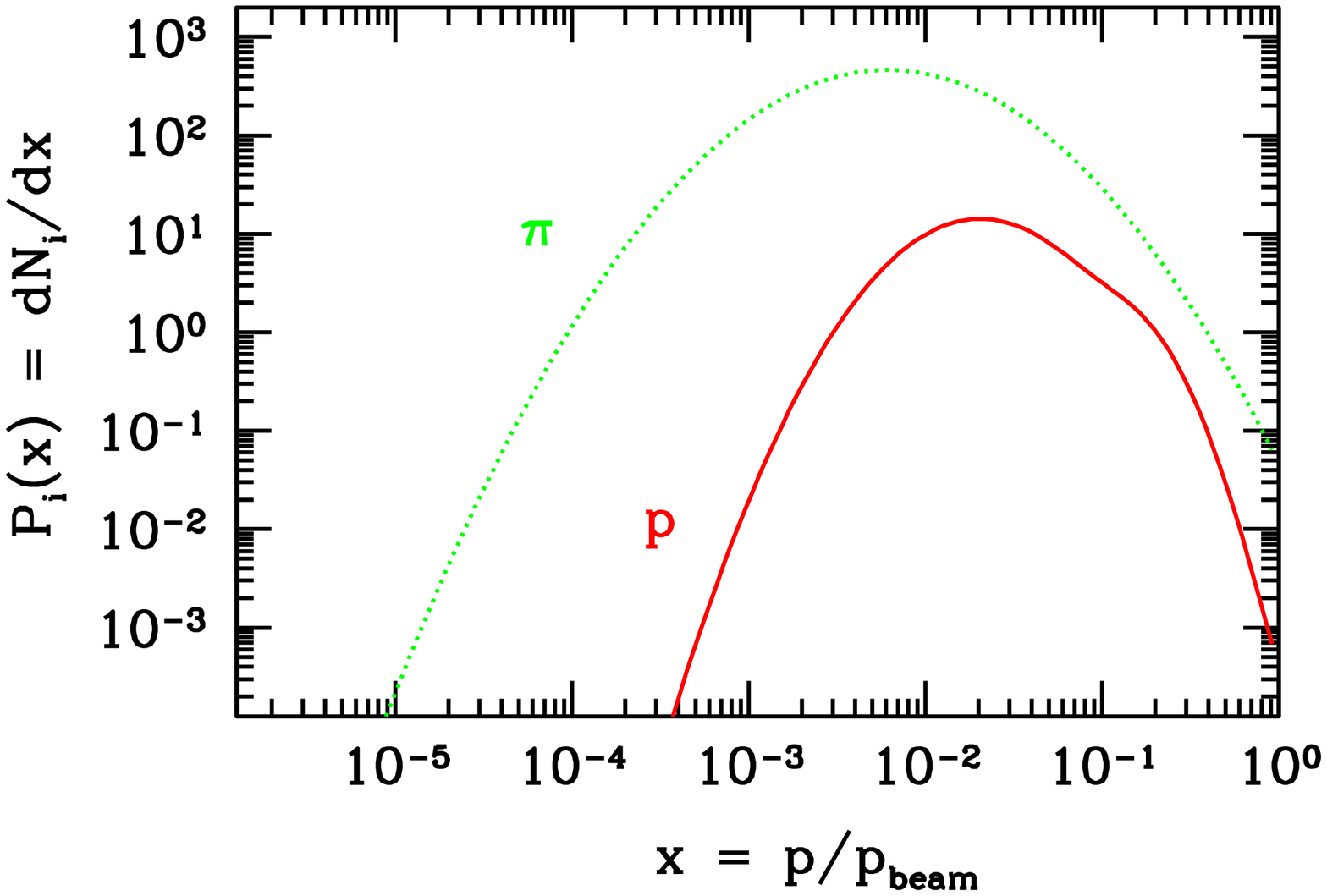}
\includegraphics[bbllx=20pt,bblly=225pt,bburx=570pt,bbury=608pt,width=7.7cm]
{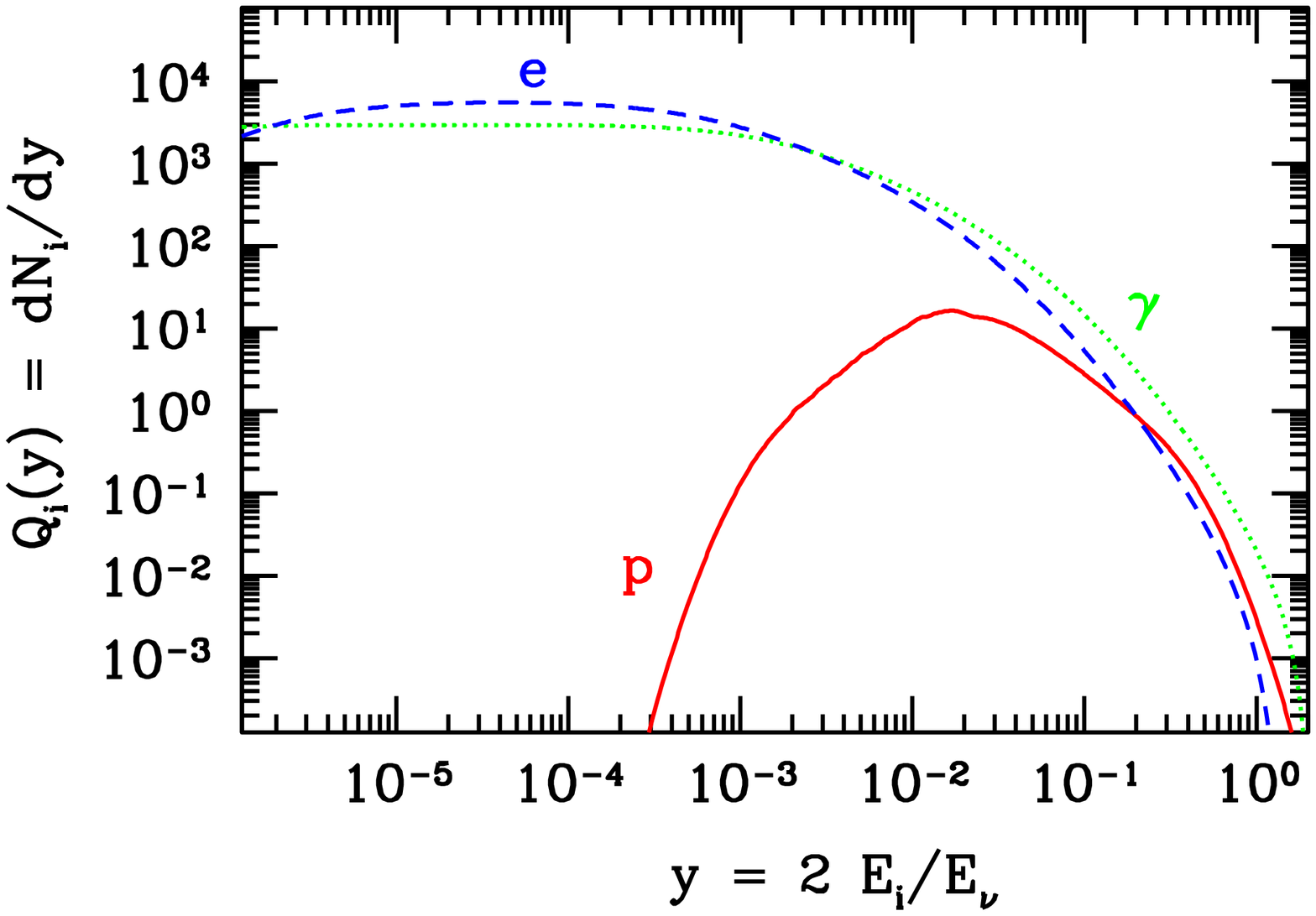}
\caption[...]{\label{prot-mom-dist}
Momentum distributions in hadronic Z decays. 
{\em Left:} Combined data from collider experiments
on proton (plus antiproton) momentum distribution (solid), 
normalized to $\langle N_p\rangle =1.04$, and charged pion momentum 
distributions (dotted), 
normalized to $\langle N_{\pi^\pm}\rangle = 16.99$.
{\em Right:} Distribution of protons (``p''; solid), photons (``$\gamma$''; 
dotted) and 
electrons (``e''; dashed) in the lab system, in which the target 
relic neutrino is at rest. 
}
\end{center}
\end{figure}

The next ingredient is the energy distribution of the produced 
protons and photons in Z decay. We combined 
existing published and some improved unpublished data on the momentum 
distribution, 
\begin{equation}
{\cal P}_p\,(x)\equiv \frac{{\rm d}N_p}{{\rm d}x}\,,
\hspace{6ex}
x\equiv \frac{p_p}{p_{\rm beam}}\,,
\end{equation}
of protons ($p$) (plus antiprotons ($\bar p$)) in Z 
decays~\cite{Akers:1994ez,Abreu:1995cu,Buskulic:1994ft,Abe:1999qh,Opal:unp}, 
see Fig.~\ref{prot-mom-dist} (left). 
The experimental data, ranging down to $x\approx 8\cdot 10^{-3}$, were 
combined with the predictions from the modified leading logarithmic 
approximation (MLLA)~\cite{Azimov:1985np}
at low $x$.
The $p+{\bar p}$ multiplicity is 
$\langle N_p\rangle = \int_0^1 {\rm d}x\, {\cal P}_p(x) = 1.04\pm 0.04$ in the 
hadronic channel~\cite{Groom:2000in}. 

In the cm system of the Z production the
angular distribution of the hadrons is determined by the
spin $1/2$ of the primary quarks and thus proportional to 
$1+w^2=1+\cos^2\theta$ (here $\theta$ is the angle between the
incoming neutrinos and the outgoing hadrons~\cite{Schmelling:1995py}).  
The energy distribution of the produced protons with energy 
$E_p$ entering the Z-burst spectrum~(\ref{p-flux}),
\begin{equation}
\label{Q-def}
\frac{{\rm d} N_{p}}{{\rm d} E_p} = 
\frac{2}{E_\nu}\,\frac{{\rm d} N_p}{{\rm d} y}
\equiv
\frac{2}{E_\nu}\,{\cal Q}_p(y)\,,
\end{equation}
with $y=2E_p/E_\nu$,
is finally obtained after a Lorentz transformation from the cm system
to the lab system,
\begin{eqnarray}
\label{Q-dist}
&&{\mathcal Q}_p(y)= \sum_{+,-} {3 \over 8} \int_{-1}^{+1} {\rm d}w\, (1+w^2)  
\\
&&\times\, {1 \over 1-w^2}
\left|
{{ \pm y-w\sqrt{y^2-(1-w^2)(2m_p/M_Z)^2}}
\over \sqrt{y^2-(1-w^2)(2m_p/M_Z)^2}}
\right|
\nonumber\\
&&{\cal P}_p\left( \frac{-wy \pm \sqrt{y^2-(1-w^2)(2m_p/M_Z)^2}}{1-w^2} 
\right), 
\nonumber
\end{eqnarray}
where $m_p$ is the proton mass.  
The first line comes from the angular distribution, 
the second line is the Jacobian and
the third one is the momentum distribution at the inverted momentum.
The scaled energy distribution ${\cal Q}_p$, as defined in Eq.~(\ref{Q-def}) 
and given by Eq.~(\ref{Q-dist}), is 
displayed in Fig.~\ref{prot-mom-dist} (right). 

Neutrons produced in Z decays will decay and end up as UHECR protons. 
They were taken into account according to 
\begin{equation}
{\mathcal Q}_{p+n}(y)=
\left(1+\frac{\langle N_n\rangle}{\langle N_p\rangle}\right)\,{\cal Q}_p(y)\,,
\end{equation}
where the neutron ($n$) (plus antineutron ($\bar n$)) multiplicity, 
$\langle N_n\rangle = \int_0^2 {\rm d}y\, {\cal Q}_n(y)$, is $\approx 4\%$ 
smaller than the proton's. 

Photons are produced in hadronic Z decays via fragmentation into neutral pions, 
$Z\to \pi^0 + X\to 2\,\gamma + X$ (cf. Fig.~\ref{illu}). 
The corresponding scaled energy distribution in the lab system, defined 
analogously to 
Eq.~(\ref{Q-def}), reads
\begin{equation}
\label{Q-dist-gamma}
{\mathcal Q}_\gamma (y)
=\int_{-1}^{1} {\rm d}w\,  
\frac{2}{1+w}\,
{\mathcal Q}_{\pi^0} 
\left( \frac{2\,y}{1+w} \right) 
\,,
\end{equation}
where the scaled energy distribution ${\mathcal Q}_{\pi^0}(y)$, with 
$y=2\,E_{\pi^0}/E_\nu$, 
is given by Eq.~(\ref{Q-dist}), with $m_p\to m_{\pi^0}$ and  ${\mathcal P}_p\to 
{\mathcal P}_{\pi^0}$, 
the momentum distribution of pions in hadronic Z decay. For the latter 
distribution, we took the measured one
of charged pions ${\mathcal P}_{\pi^\pm}$ from hadronic Z 
decay~\cite{Akers:1994ez,Abreu:1995cu,Buskulic:1994ft,Abe:1999qh,Opal:unp}
(cf. Fig.~\ref{prot-mom-dist} (left)), 
normalized such that $\langle N_\gamma\rangle = \int_0^2 {\rm d}y\, {\cal 
Q}_\gamma (y)= 20.97$~\cite{Groom:2000in}.  

Electrons (and positrons) from hadronic Z decay are also relevant for the 
development of electromagnetic cascades.
They stem from decays of secondary charged pions, $Z\to \pi^\pm +X\to e^\pm + 
X$ (cf. Fig.~\ref{illu}), 
and their scaled energy distribution in terms of $y=2\,E_e/E_\nu$ reads 
\begin{eqnarray}
{\mathcal Q}_{e^\pm}(y) &=&
\int_{-1}^{1}dw \int_0^2 dx 
\frac{1}{xw+\sqrt{x^2+(2\,m_e/m_\pi)^2}}
\\[1ex] \nonumber && \times
{\mathcal Q}_{\pi^\pm}
\left( 
\frac{2\, y}{xw+\sqrt{x^2+(2\,m_e/m_\pi)^2}} \right) 
{\mathcal P}_e (x)
\,,
\end{eqnarray}
where ${\mathcal P}_e$ is the momentum distribution of the electrons in the 
rest system of the 
charged pion. The energy distribution ${\mathcal Q}_{e^\pm}$ is also 
displayed in Fig.~\ref{prot-mom-dist} (right).

\subsection{\label{prop}Propagation of nucleons and photons} 

The cosmic microwave background 
is known to a high accuracy.  It plays the key role in the 
determination of the probability $P_p(r,E_p;E)$
that a proton created at a distance $r$ with energy $E_p$ arrives
at Earth above the threshold energy $E$, suggested in 
Ref.~\cite{Bahcall:2000ap} and
determined for a wide range of parameters in Ref.~\cite{Fodor:2001yi}. 
The propagation function $P_p$ takes into 
account the fact that protons of extragalactic (EG) origin and energies above 
$\approx 4\cdot 10^{19}$ eV 
lose a large fraction of their energy
due to pion and $e^+e^-$ production 
through scattering on the CMB and due to their 
redshift~\cite{Greisen:1966jv,Zatsepin:1966jv}.
In our analysis we went, according to 
\begin{equation}
\label{z-H}
{\rm d}z = - (1+z)\,H(z)\,{\rm d}r/c\,,
\end{equation}
out to distances $R_{\rm max}$ (cf.~(\ref{p-flux})) 
corresponding to redshift $z_{\rm max} = 2$. 
We used the expression 
\begin{equation}
\label{H-Omega}
H^2(z) = H_0^2\,\left[ \Omega_{M}\,(1+z)^3 
+ \Omega_{\Lambda}\right]
\end{equation} 
for the relation of the Hubble expansion rate at redshift $z$ to the present 
one.
Uncertainties of the latter, $H_0=h$ 100 km/s/Mpc, with 
$h=(0.71\pm 0.07)\times^{1.15}_{0.95}$~\cite{Groom:2000in}, 
were included. 
In Eq.~(\ref{H-Omega}), $\Omega_{M}$ and $\Omega_{\Lambda}$, with 
$\Omega_M+\Omega_\Lambda =1$, are the present 
matter and vacuum energy densities in terms of the critical density. 
As default values we chose
$\Omega_M = 0.3$ and $\Omega_\Lambda = 0.7$, as favored today. Our results
turn out to be pretty insensitive to the precise values of the cosmological 
parameters.

\begin{figure}
\begin{center}
\includegraphics[bbllx=20pt,bblly=221pt,bburx=570pt,bbury=608pt,width=7.7cm]
{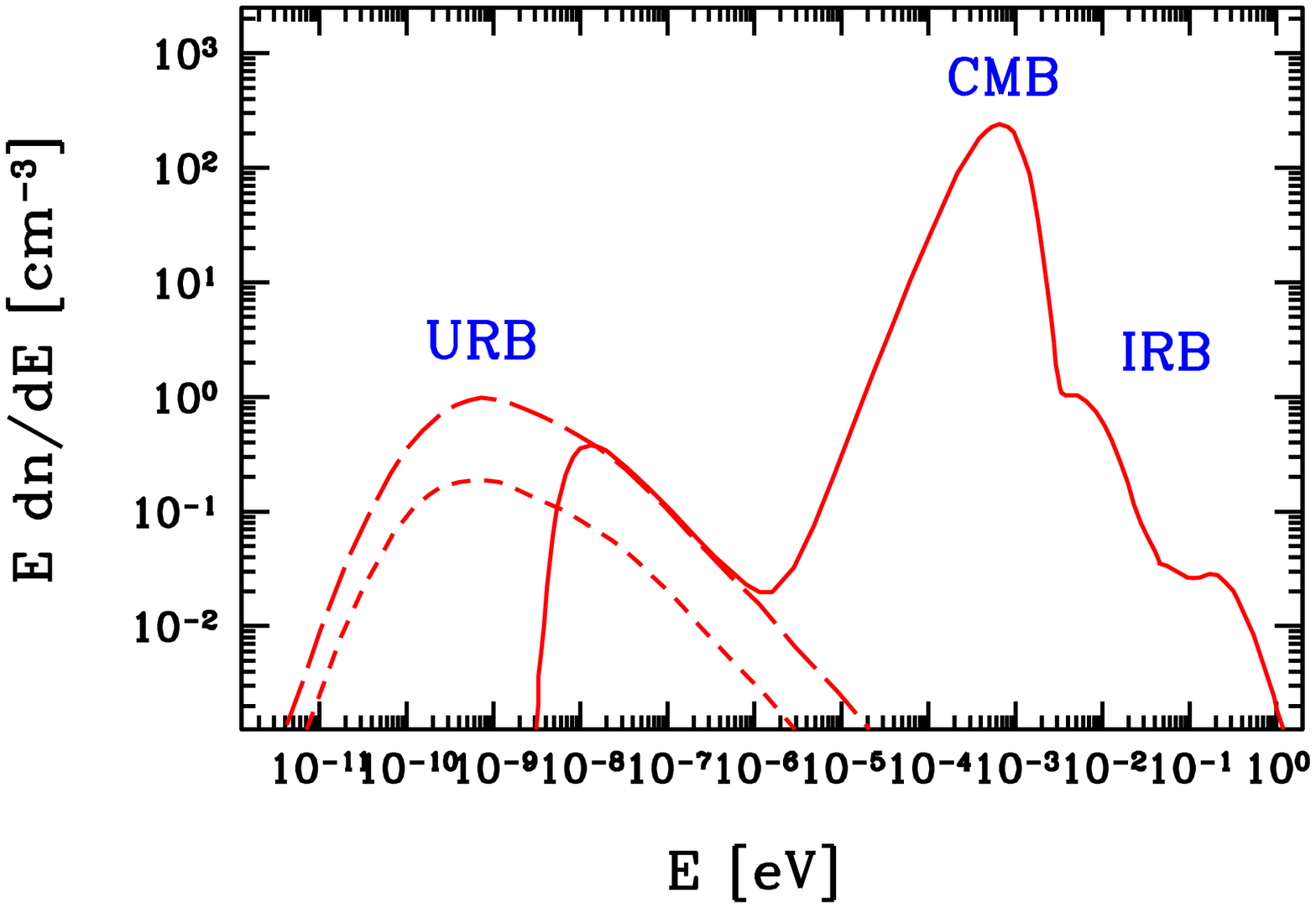}
\includegraphics[bbllx=20pt,bblly=221pt,bburx=570pt,bbury=608pt,width=7.7cm]
{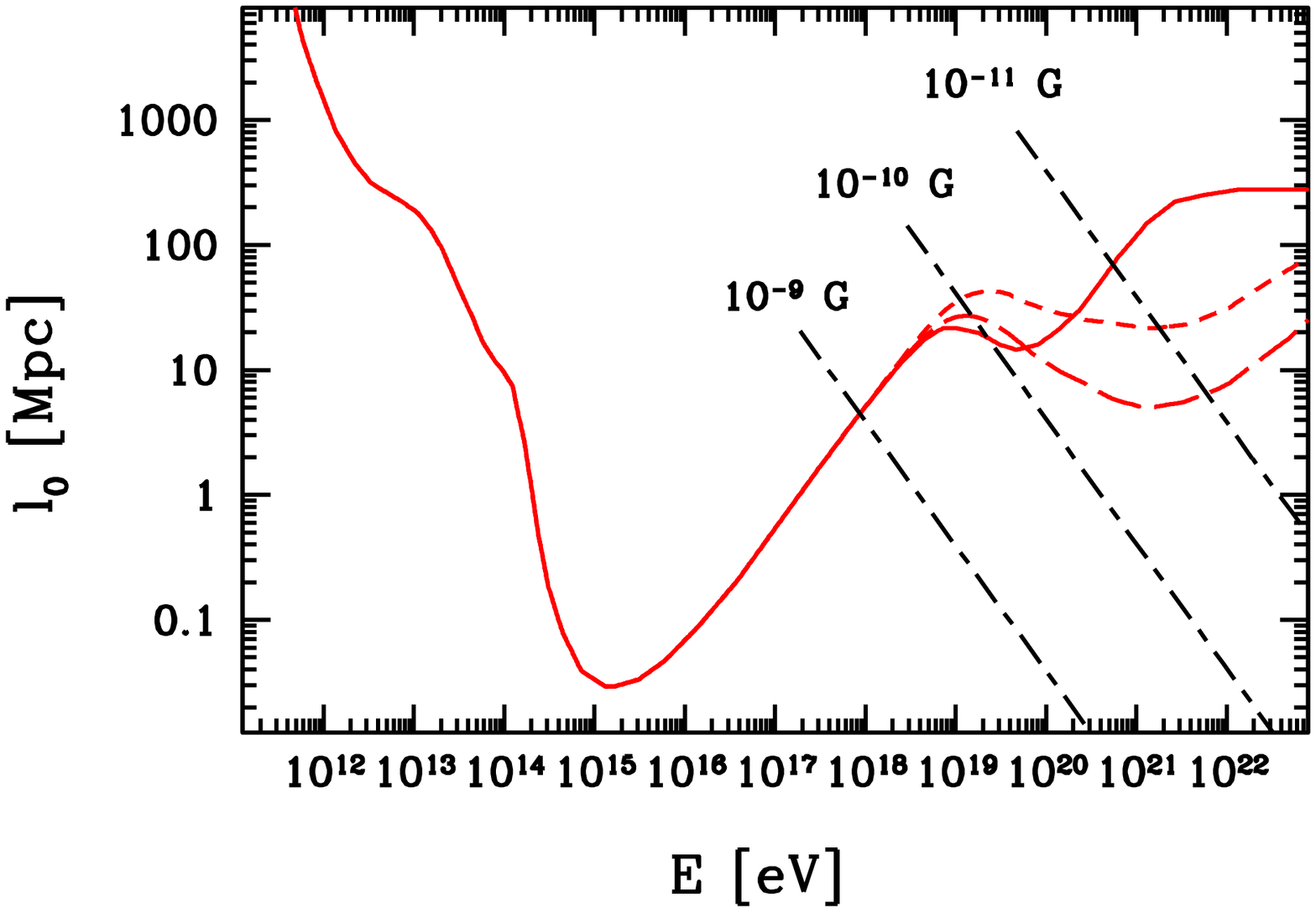}
\caption[...]{\label{ph_mfp}
{\em Left:} The intensity spectrum of the diffuse extragalactic photon 
background 
at redshift $z=0$ (solid). 
Different estimates of the universal radio 
background (URB) are also indicated: 
high URB (long-dashed) and moderate URB (short-dashed). 
{\em Right:} Photon energy attenuation length $l_0$ at $z=0$ corresponding to 
the photon 
background shown on the left (solid).
Variations of $l_0$ arising from different assumptions about the URB 
are also indicated: high URB (long-dashed) and moderate URB (short-dashed).
The energy attenuation 
length for electrons due to synchrotron radiation for different 
magnitudes of the extragalactic magnetic fields is also shown 
(long-dashed-short-dashed).
}
\end{center}
\end{figure}

We determined $P_p(r,E_p;E)$ in a range
of $r\leq 4000$~Mpc, $10^{18}$~eV $\leq E_p\leq 10^{26}$~eV, and $10^{18}$~eV 
$\leq E\leq 10^{26}$~eV, 
for several  fixed values of the cosmological parameters.
The simulation was carried out in small ($10$~kpc) steps in $r$. For each step, 
the statistical energy losses due to pion/$e^+e^-$ production and redshift were 
taken 
into account~\cite{Fodor:2001yi}. 
In this connection, the advantage of our formulation of the Z-burst spectrum 
in terms of the probability  $P_p(r,E_p;E)$ becomes evident. We have to 
determine the latter       
only once and for all. Without the use of $P_p(r,E_p;E)$, we would have to 
perform a simulation
for any variation of the input spectrum, notably for any change in the neutrino 
mass. 
Since $P_p(r,E_p;E)$ is of universal usage, we made the 
corresponding  
numerical data for the probability distribution 
$(-)\partial P_p (r,E_p;E)/\partial E$   
available for the public via the World-Wide-Web 
URL {\it http://www.desy.de/\~{}uhecr}.  

The determination of the photon propagation function $P_\gamma (r,E_\gamma ;E)$,
entering the photon flux prediction, was done as follows. 
In distinction to the case of the proton propagation function, we used here 
the continuous energy loss  (CEL) approximation which 
largely simplifies the work and reduces the required computer resources.
In the CEL 
approximation, the energy (and number) of the detected photons is a unique 
function of
the initial energy and distance, and statistical fluctuations are neglected.
A full simulation of the photon propagation function will be the subject of a 
later
work.   

The processes that were taken into account are pair production on the 
diffuse extragalactic 
photon background (cf. Fig.~\ref{ph_mfp} (left)), double pair production 
and inverse Compton scattering of
the produced pairs. We comment also on synchrotron radiation in a possible 
extragalactic magnetic field (EGMF).  
For the energy attenuation length of the photons due to these processes, we 
exploited the values quoted in  
Ref.~\cite{Lee:1998fp} (see also Ref.~\cite{Protheroe:1996ft}) and the further 
ones presented in
Fig.~\ref{ph_mfp} (right) which incorporate various assumptions
about the poorely known universal radio (URB) (from 
Ref.~\cite{Protheroe:1996si}) and infrared (IRB) backgrounds. 
We shall analyse later the dependence of the neutrino mass and other fit 
parameters 
on these variations. Note, that, in view of the recent URB estimates in 
Ref.~\cite{Protheroe:1996si}, the ones presented in Fig.~\ref{ph_mfp} (left)), 
which are based on Ref.~\cite{Clark:1970}, can be referred to as ``minimal'' 
URB.  

The computation of the photon propagation function $P_\gamma (r,E_\gamma ;E)$ 
was carried out in the following way.
The energy attenuation of photons in the CEL approximation 
was calculated according to 
\begin{eqnarray}
\label{E-atten}
dE=-E\left(\frac{{\rm d}r}{l_z(E)}-\frac{{\rm d}z}{1+z}\right)\,,
\end{eqnarray}
where $l_z(E)=(1+z)^{-3}\, l_0(E(1+z))$ is the energy attenuation length at 
redshift $z$.
The number of photons was assumed to be constant at ultrahigh energies $\gwig 
10^{18}$~eV, 
due to the small inelasticities in this energy range. Below, it was increased 
in a way to 
maintain energy conservation (except for the redshift contribution):
\begin{eqnarray}
\label{n-phot}
{\rm d}N_\gamma =-N_\gamma\, \frac{{\rm d}r}{l_z(E)}\,.
\end{eqnarray}
The $P_\gamma (r,E_\gamma ;E)$ function was then obtained by integration of 
these equations.
In the ultrahigh energy region -- which is most relevant for us since we 
performed our fit to the 
cosmic ray data there --  
the approximation described above gives the photon flux quite reliable, while 
at lower energies 
it yields an upper bound.

\subsection{\label{fluxes}UHEC\boldmath$\nu$\unboldmath\ fluxes}

\begin{figure}
\begin{center}
\vspace*{2.0mm} 
\includegraphics[bbllx=20pt,bblly=221pt,bburx=570pt,bbury=608pt,width=8.0cm]
{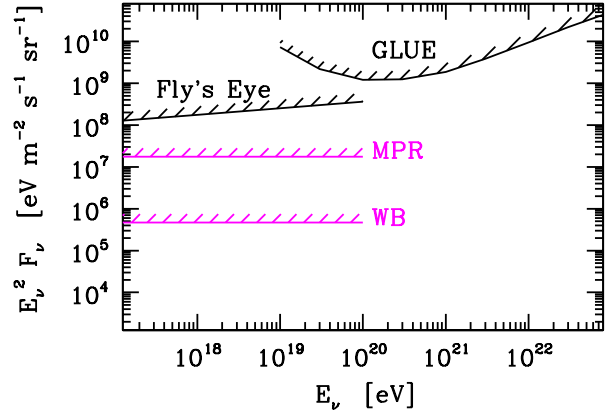}
\caption{\label{flux_upp_lim}
Upper limits on differential neutrino fluxes in the ultrahigh energy regime.
Shown are experimental upper limits 
on $F_{\nu_e}+F_{\bar\nu_e}$ from 
Fly's Eye and                    
on $\sum_{\alpha =e,\mu} (F_{\nu_\alpha }+F_{\bar\nu_\alpha })$ from 
the Gold\-stone lunar ultrahigh energy neutrino ex\-pe\-ri\-ment 
GLUE, as well as theoretical upper limits on 
$F_{\nu_\mu}+F_{\bar\nu_\mu}$
from ``visible'' (``WB'') and 
``hidden'' (``MPR'') hadronic astrophysical sources.
}
\end{center}
\end{figure}

The differential fluxes $F_{\nu_i}$ of UHE$\nu$s are unknown. 
Present experimental upper limits on these fluxes are rather poor.
The results of the Fly's Eye~\cite{Baltrusaitis:1985mt} and 
Gold\-stone lunar ultrahigh energy (GLUE)~\cite{Gorham:2001aj} neutrino 
experiments are summarized of Fig.~\ref{flux_upp_lim}.

What are the theoretical expectations for diffuse UHEC$\nu$ fluxes? 
More or less guaranteed are the so-called 
cosmogenic neutrinos which are produced when ultrahigh energy cosmic protons
scatter inelastically off the cosmic microwave background radiation
in processes such as $p\gamma\to \Delta\to n\pi^+$,
where the produced pions subsequently 
decay.
These fluxes (for recent estimates, see 
Refs.~\cite{Yoshida:1993pt,Protheroe:1996ft,Yoshida:1997ie,Engel:2001hd,%
Kalashev:2002kx}) 
represent reasonable 
lower limits, 
but turn out to be insufficient for the
Z-burst scenario. 
Recently, theoretical upper limits on the ultrahigh energy cosmic neutrino flux 
have been given 
in Refs.~\cite{Waxman:1999yy,Bahcall:2001yr,Mannheim:2001wp}. Per construction, 
the upper
limit from ``visible'' hadronic astrophysical sources, i.\,e. from those 
sources which are transparent to ultrahigh 
energy cosmic protons
and neutrons, is of the order of the cosmogenic neutrino flux
and shown in Fig.~\ref{flux_upp_lim} (``WB''; 
cf.~Refs.~\cite{Waxman:1999yy,Bahcall:2001yr}). 
Also shown in this figure (``MPR'') is 
the much larger upper limit from ``hidden'' hadronic astrophysical sources, 
i.\,e. from those sources from which only 
photons and neutrinos can escape~\cite{Mannheim:2001wp}. 
A recent detailed simulation showed that these bounds can be exceeded
if the spectrum of the source is hard enough~\cite{Kalashev:2002kx}.
Even larger fluxes at ultrahigh energies may arise if the hadronic 
astrophysical sources emit
photons only in the sub-MeV region -- thus evading the ``MPR'' bound in 
Fig.~\ref{flux_upp_lim} -- 
or if the neutrinos are produced via the decay of superheavy 
relic particles.

In this situation of insufficient knowledge, we took the following approach 
concerning the flux of 
ultrahigh energy cosmic neutrinos, $F_{\nu_i}(E_{\nu_i},r)$. It is assumed to 
have the form 
\begin{equation}
F_{\nu_i}(E_{\nu_i},r)=F_{\nu_i}(E_{\nu_i},0)\,(1+z)^\alpha\,,
\end{equation} 
where $z$ is the redshift and where $\alpha$ characterizes the cosmological 
source evolution  
(see also Refs.~\cite{Yoshida:1997ie,Yoshida:1998it,Kalashev:2001sh}). The flux 
at zero redshift, 
$F_{\nu_i}(E_{\nu_i},0)\equiv F_{\nu_i}(E_{\nu_i})$,
is left open. 
For hadronic astrophysical sources it is expected to fall off power-like, 
$F_{\nu_i}(E_{\nu_i})\propto E_{\nu_i}^{-\gamma}$, 
$\gamma\gwig 1$, at high energies. Due to this fact and because of the strong 
resonance peaks in 
the $\nu_i\bar\nu_i$ annihilation 
cross section~(\ref{z-res-cs-roulet}) at the resonance energies~(\ref{eres}), 
the Z-burst rate will be 
only sensitive to 
the flux at the resonant energy of
the heaviest neutrino. Of course, the latter may be nearly degenerate with the 
other neutrino 
mass eigenstates, 
$m_{\nu_i}\approx m_\nu$, as it is the case for $m_{\nu_3}\,\gwig\, 0.1$~eV in 
a three flavour scenario. 
Correspondingly, our fit to the UHECR data is sensitive only to 
\begin{equation}
\label{fnures}
F_\nu^{\rm res} = \sum_i \left[ F_{\nu_i}(E_{\nu_i}^{\rm 
res})+F_{\bar\nu_i}(E_{\nu_i}^{\rm res})
\right] \,,
\end{equation}
where the sum extends over the number of mass eigenstates which are 
quasi-degenerate with the heaviest neutrino.
Note, finally, that, independently of the production
mechanism, neutrino oscillations result in a uniform $F_{\nu_i}$ mixture for 
the 
different mass eigenstates $i$.   

\begin{figure} \begin{center}
\hspace*{-2.cm}
\vspace*{0.8cm}
\includegraphics[angle=90,width=8.7cm]
{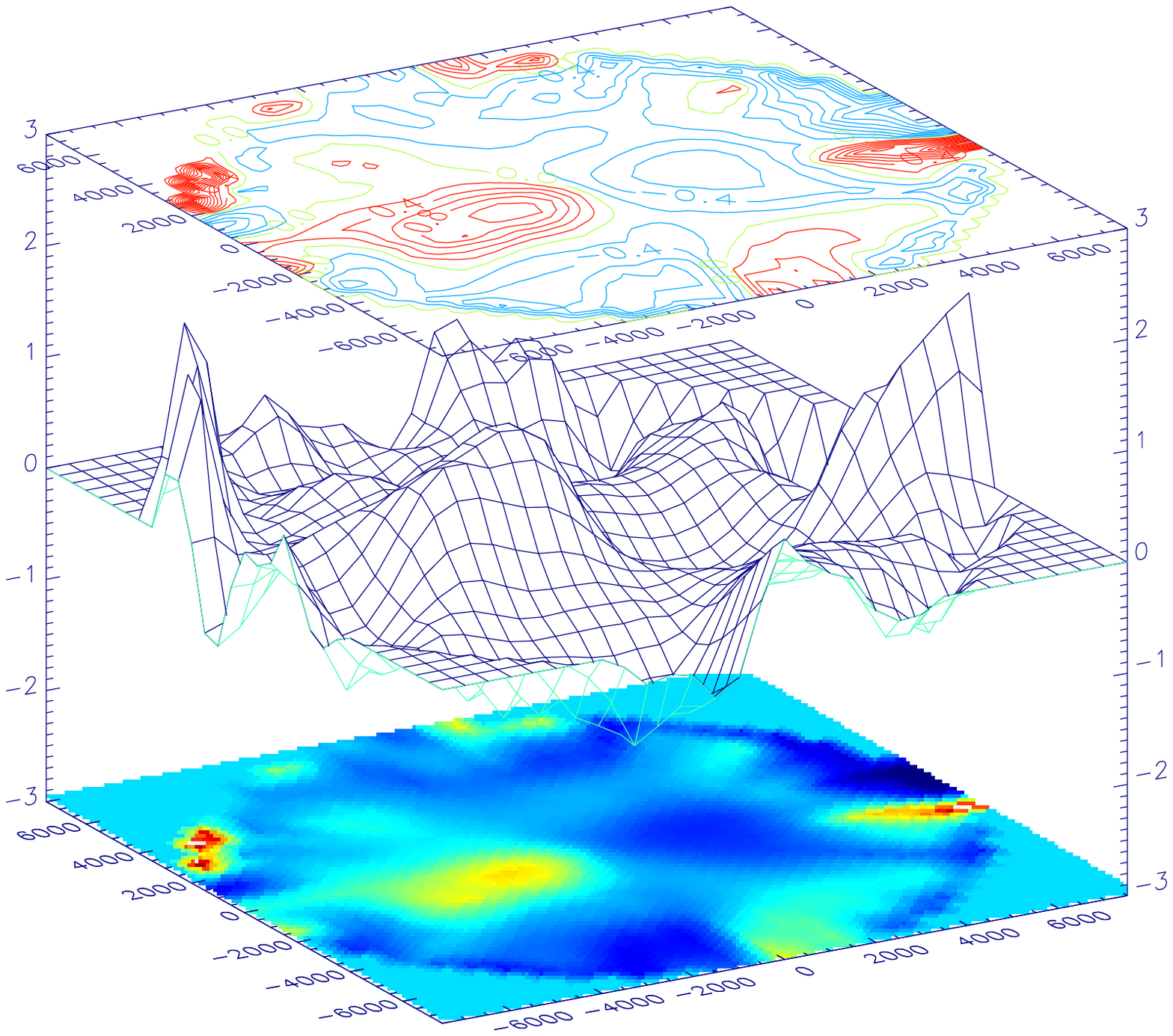}
\vspace*{-0.8cm}
\includegraphics[bbllx=20pt,bblly=225pt,bburx=570pt,bbury=608pt,width=7.7cm]
{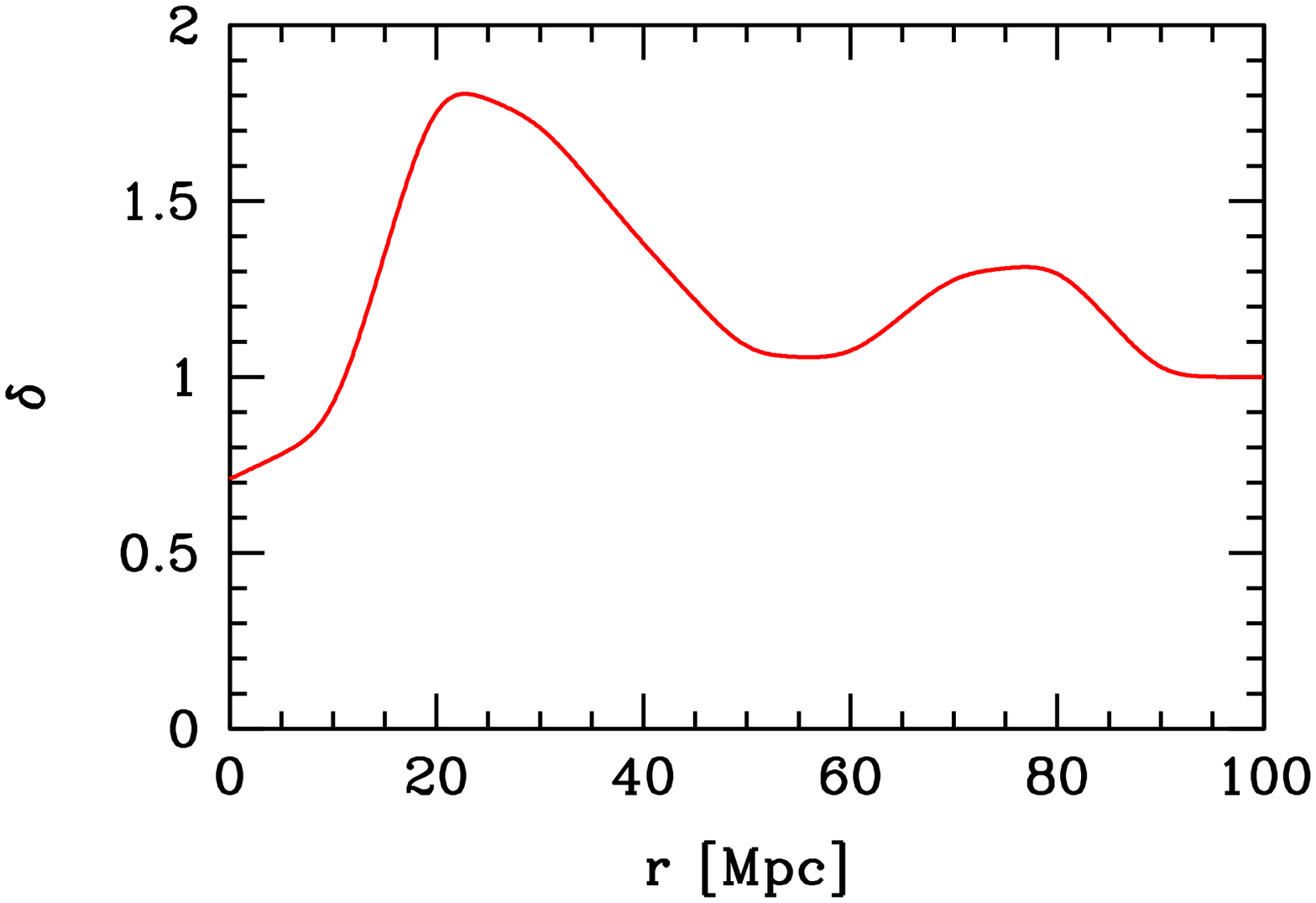}
\caption[...]{\label{dens-prof}
{\em Left:}
Mass density fluctuation field $\delta$ along the 
supergalactic plane as obtained from
peculiar velocity measurements. Shown are contours
in intervals of $\delta =0.2$, surface maps on a grid of spacing 
$500$ km\,s$^{-1}$, corresponding to $5\,h^{-1}$ Mpc, with the height 
proportional to
$\delta$, and contrast maps. One recognizes some well-known structures in the
nearby volume such as the Great Attractor at supergalactic coordinates
(SGX $\sim -2000$ km\,s$^{-1}$, SGY $\sim -500$ km\,s$^{-1}$), 
the Perseus-Pisces complex 
(SGX $\sim 6000$ km\,s$^{-1}$, SGY $\sim -1000$ km\,s$^{-1}$), 
and the large void
(SGX $\sim 2500$ km\,s$^{-1}$, SGY $\sim 0$ km\,s$^{-1}$) in between.
{\em Right:} 
Mass density fluctuation field obtained from above data,  
averaged over all directions, for $h=0.71$. The overdensities at around 20 and 
80 Mpc 
reflect the Great Attractor and the Perseus-Pisces complex, respectively. 
}
\end{center}
\end{figure}

\subsection{\label{nunumb}Neutrino number density}

The dependence of the relic neutrino number density $n_{\nu_i}$ on the distance 
$r$ is the last ingredient of the Z-burst spectra. 

The main question is whether there is remarkable clustering of the relic neutrinos 
within
the local GZK zone of about 50 Mpc. It is known that the 
density distribution of relic neutrinos as hot dark matter follows the total 
mass 
distribution; however, with  less clustering~\cite{Ma:1998aw,Primack:2000iq,%
Singh:2002de}.
To take this into account, the shape of the $n_{\nu_i}(r)$ 
distribution was varied, for distances below 100 Mpc,  
between the standard cosmological homogeneous case 
and that of 
the total mass distribution obtained from peculiar velocity 
measurements~\cite{daCosta:1996nt} (cf. Fig.~\ref{dens-prof} (left)). 
These peculiar measurements suggest relative overdensities of at most a factor 
$f_\nu =2\div 3$, 
depending on the grid spacing (cf. Fig.~\ref{dens-prof} (right)). 
A relative overdensity $f_\nu = 10^2\div 10^4$ in our neighbourhood, as it was 
assumed in earlier 
investigations of the Z-burst 
hypothesis~\cite{Fargion:1999ft,Weiler:1999sh,Waxman:1998yh,Yoshida:1998it,%
Blanco-Pillado:2000yb,McKellar:2001hk}, 
seems unlikely in view of these data. 
Our quantitative results turned out to be rather insensitive to the 
variations of the overdensities within the considered range.
For scales larger than 100 Mpc the relic neutrino 
density was taken according to the big bang cosmology prediction, 
$n_{\nu_i}=56\cdot (1+z)^3$ cm$^{-3}$.

\section{\label{sect:determination}Determination of the required neutrino mass 
and the necessary UHEC$\nu$ flux}

Our comparison with the observed spectrum included published UHECR data
of AGASA~\cite{Takeda:1998ps}, Fly's Eye~\cite{Bird:yi,Bird:wp,Bird:1994uy}, 
Haverah Park~\cite{Lawrence:cc,Ave:2000nd}, 
and HiRes~\cite{Kieda00}, as well as unpublished one from the World Wide Web 
pages of 
the experiments on 17/03/01.
Due to normalization difficulties we did not use the Yakutsk~\cite{Efimov91} 
results. The latest HiRes data~\cite{Abu-Zayyad:2002sf} are not yet included.
We shall take into account the fact that above $4\cdot 10^{19}$~eV less than 
$50\,\%$ of 
the cosmic rays can be photons at the $95\,\%$ confidence level 
(C.L.)~\cite{Ave:2001xn} 

As usual, each logarithmic unit between 
$\log (E/\mbox{eV})=18$ and $\log (E/\mbox{eV})=26$
is divided into ten bins.
The predicted number of UHECR events in a bin is given by
\begin{equation}
\label{flux}
 N(i)= {\mathcal E}
\int_{E_i}^{E_{i+1}}
{\rm d}E
\left[
F_{p|{\rm bkd}}(E  )
+ F_{p(+\gamma )|Z} (E )\right],
\end{equation}
where ${\mathcal E} \approx 8\,\cdot 10^{16}$ m$^2\cdot$\,s\,$\cdot$\,sr
is the total exposure of the experiments (estimated from the highest energy
events and the corresponding fluxes) and where 
$E_i=10^{(18+i/10)}$~eV is the lower bound of the $i^{\rm th}$ energy bin. The 
first 
term in Eq.~(\ref{flux}), $F_{p|{\rm bkd}}$, corresponds to the diffuse
background of ordinary cosmic rays from unresolved astrophysical sources. 
Below the GZK cutoff, it should have the usual and experimentally observed 
power-law form~\cite{Nagano:2000ve}.   
The second term represents the sum of the proton and photon spectra, 
$F_{p|Z}+F_{\gamma |Z}$ from Z-bursts.

The separation of the flux into two terms (one from the power-law   
background and one from the Z-burst) is physically well motivated. The
power-law part below the GZK cutoff is confirmed experimentally. For
extragalactic sources, it should suffer from the GZK effect. 
In the Z-burst scenario, cosmic rays are coming from another independent source
(Z-bursts), too. What we observe is the sum of the two. As the detailed
fits in the next section will show, the flux from Z-bursts is much smaller in 
the low energy
region than the flux of the power-law background. Correspondingly, the low 
energy part
of the spectrum (between $10^{18.5}$ and $10^{19.3}$~eV) has very little 
influence on the
Z-burst fit parameters, notably on the neutrino mass.
We investigated three possibilities for the background term $F_{p|{\rm bkd}}$ . 

\begin{figure}
\includegraphics[bbllx=20pt,bblly=221pt,bburx=570pt,bbury=608pt,width=7.9cm]
{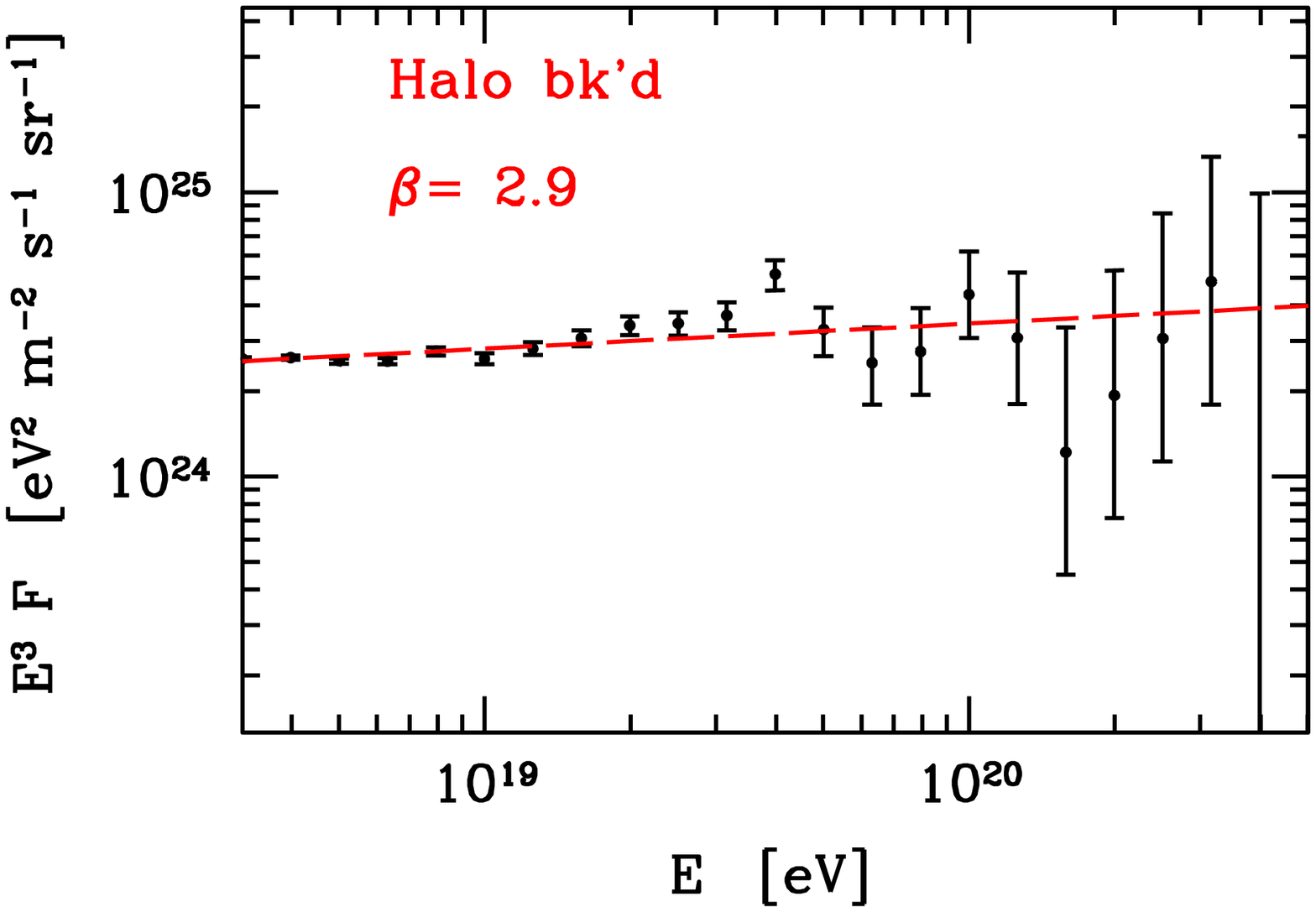}
\includegraphics[bbllx=20pt,bblly=221pt,bburx=570pt,bbury=608pt,width=7.9cm]
{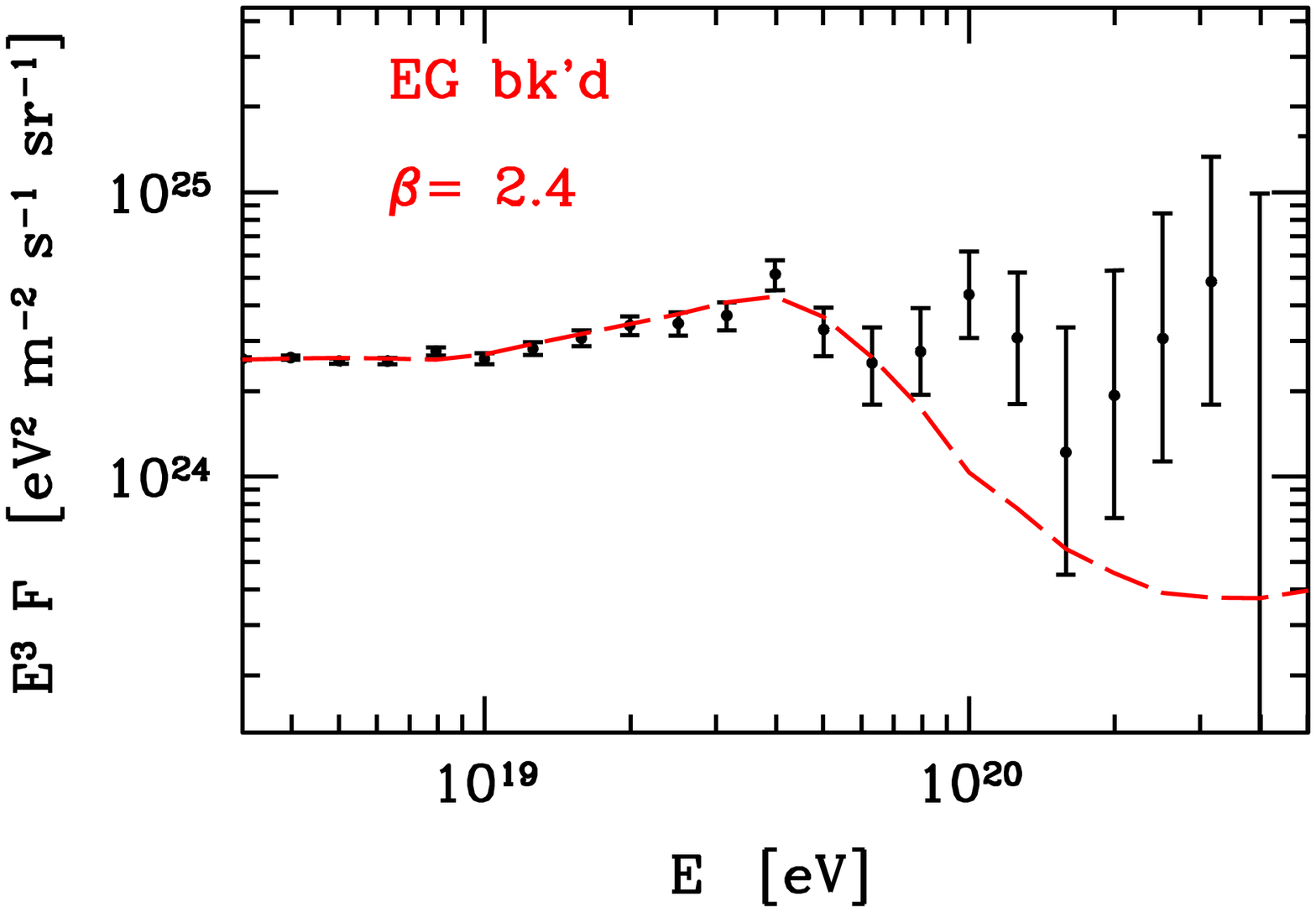}
\caption[...]{\label{fit_normal}
The available UHECR data with their error bars
and the best fits (long-dashed) from ordinary cosmic ray protons originating 
from our local neighbourhood
(``halo background''; {\em left}) and originating from diffuse extragalactic 
sources 
(``EG background''; {\em right}), respectively.
For the latter case, the bump at $4\cdot 10^{19}$~eV represents protons 
injected at
high energies and accumulated just above the GZK cutoff due to their energy
losses. The predicted fall-off for energies above $4\cdot 10^{19}$~eV can be 
observed.
}
\end{figure}

The first possibility is to assume that the ordinary cosmic rays originate from
our galactic halo (or at least from within the GZK zone of about 50~Mpc). 
This scenario will be referred to as ``halo background'' in the following.
We should note, however, that there are no known astrophysical sources in
the arrival directions of the highest energy events. This fact makes this 
possibility unlikely.
In the halo background case the spectrum of ordinary cosmic rays is not 
distorted by propagation effects, so we can assume the usual power-law
behaviour~\cite{Takeda:1998ps,Nagano:2000ve} (cf. Fig.~\ref{fit_normal} (left)), 
\begin{eqnarray}
\label{pow-law-halo}
{\mathcal E}\,F_{p|{\rm bkd}}(E;A,\beta ) = 
\frac{A}{1\,{\rm eV}}\,
\left( \frac{E}{1\,{\rm eV}}\right)^{-\beta}
 \hspace{3ex} ({\rm Halo\ bk'd})
\,.
\end{eqnarray}

The second, most plausible possibility is to   
assume that the ordinary cosmic rays are 
protons from 
uniformly distributed, extragalactic sources -- thus we call it   
``extragalactic background''.
In view of the observed, practically uniform distribution of the 
arrival directions of UHECRs, this 
assumption seems to be 
phenomenologically more realistic than the halo background model. 

The extragalactic background suffers of course from GZK attenuation. 
Correspondingly, we take the above 
power-law~(\ref{pow-law-halo}),  
$A \cdot E^{-\beta}_p$, as an injection spectrum and determine the final
spectrum using the proton propagation function $P_p(r,E_p;E)$, 
\begin{eqnarray}
{\mathcal E}\,F_{p|{\rm bkd}}(E;A,\beta ) &= 
 \int_0^\infty {\rm d}E_p \int_0^{R_{\rm max}} {\rm d}r 
\,\left(1+z(r)\right)^3
\\[1ex] \label{pow-law-eg} & \nonumber
\times\,
\frac{A}{1\,{\rm eV}}\,  \left( \frac{E_p}{1\,{\rm eV}}\right)^{-\beta}\,
(-)\frac{\partial P_p(r,E_p;E)}{\partial E}
 \hspace{3ex} 
({\rm EG\ bk'd})\,.
\end{eqnarray}
The predicted spectrum of the extragactic background protons shows an 
accumulation at around the GZK 
scale $4\cdot 10^{19}$~eV and a sharp drop beyond (see Fig.~\ref{fit_normal} 
(right)). 

For comparison with recent work on the Z-burst scenario~\cite{Kalashev:2001sh}, 
as a third possibility, we have  done also fits with no background component, $F_{p|{\rm bkd}}=0$ in 
Eq.~(\ref{flux}), and a larger lower end, $\log (E_{\rm min}/\mbox{eV})= 
19.4\div 20.0$.

The Z-burst part of the predicted flux~(\ref{flux}) has several parameters.
As it has already been mentioned, this flux is only sensitive to the 
mass of the heaviest neutrino, this will be denoted in the following by
$m_\nu$. The Z-burst spectra can be rewritten as
\begin{eqnarray}
{\mathcal E}\,F_{i|Z}(E;B,m_\nu ) &=
B\, \int_0^\infty {\rm d}E_i \int_0^{R_{\rm max}} {\rm d}r 
\,\left(1+z(r)\right)^{3+\alpha}\,\delta_n (r) 
\\[1ex]\label{fit-zburst-spectra}\nonumber  
 &\times
\frac{4\,m_\nu}{M_Z^2}\,
{\mathcal Q_i}\left( y = \frac{4\,m_\nu\,E_i}{M_Z^2} \right)
(-)\frac{\partial P_i(r,E_i;E)}{\partial E}
,
 \hspace{3ex} i=p,\gamma\,,
\end{eqnarray}
where 
$\alpha$ is the cosmological evolution parameter,  
$\delta_n(r)$ is the mass density fluctuation field (cf. Fig.~\ref{dens-prof} 
(right)), normalized to one,  
and $\mathcal Q_i$ are the boosted momentum distributions from hadronic Z 
decay, normalized 
to $\langle N_{p+n}\rangle = 2.04$, for $i=p$, and to $\langle N_\gamma\rangle 
= 2\,\langle N_{\pi^0}\rangle +
\langle N_{\pi^\pm}\rangle = 37$, for $i=\gamma$. 
After fixing $\alpha$ we have two fit parameters left,
the mass $m_\nu$ of the heaviest neutrino and the 
overall normalization $B$, which may be expressed, using 
Eqs.~(\ref{eresfres}) and
(\ref{fnures}), in terms of the original quantities entering the Z-burst 
spectra, as 
\begin{eqnarray}
\label{B-uhecnu-fluxes}
\frac{B}{{\mathcal E}}= {\rm Br}(Z\to {\rm hadrons})\,R_{\rm max}\,\langle 
n_{\nu_i}\rangle_0\,
\langle \sigma_{\rm ann}\rangle\,E_\nu^{\rm res}\,F_\nu^{\rm res}
\,.
\end{eqnarray}
We can see that the required UHE$\nu$ flux can easily be obtained from
the $B$ parameter.
The background term (when it is nonzero) has two parameters, the $\beta$
exponent and the $A$ normalization factor.

The expectation value for the number of events in a bin is given
by Eq.~(\ref{flux}). 
To determine the most probable value for $m_{\nu_j}$ we used the maximum 
likelihood method and minimized~\cite{Fodor:2001za} the 
$\chi^2(\beta,A,B,m_{\nu})$,
\begin{eqnarray} 
\label{chi}
\chi^2=
\sum_{\log\left(\frac{E_i}{{\rm eV}}\right)=18.5}^{\log\left(\frac{E_i}{\rm 
eV}\right)=26.0}
2\left[ N(i)-N_{\rm o}(i)+N_{\rm o}(i)
\ln\left( N_{\rm o}(i)/N(i)\right) \right],
\end{eqnarray}
where $N_{\rm o}(i)$ is the total number of observed events in the $i^{\rm th}$
bin. 
Since the Z-burst scenario results
in a quite small flux for lower energies, 
we took the lower bound just below the ``ankle'': $E_{\rm min}=10^{18.5}$~eV. 
Our results are
insensitive to the definition of the upper end (the flux is
extremely small there) for which we chose $\log (E_{\rm max}/\mbox{eV})=26$.
The uncertainties of the
measured energies are about 30\% which is one bin. 
Using a Monte Carlo analysis, we took these uncertainties into account 
and included the corresponding variations in our final error estimates.

\subsection{Fit results}

Our fitting procedure involves four parameters: $\beta,A,B$ and 
$m_\nu$. The minimum of the $\chi^2(\beta,A,B,m_\nu)$ function is 
$\chi^2_{\rm min}$ at $m_{\nu\, {\rm min}}$ which is
the most probable value for the mass, whereas
the 1\,$\sigma$ (68\%) confidence interval for $m_\nu$ 
is determined by 
\begin{equation}
\chi^2(\beta',A',B',m_\nu)\equiv \chi^2_o(m_\nu)=\chi^2_{\rm min}+1
\,.
\end{equation}
Here $\beta'$, $A'$, $B'$ are defined in such a way that the 
$\chi^2$ function is minimized in $\beta,A$
and $B$, at fixed $m_\nu$.

\begin{figure} \begin{center}
\includegraphics[bbllx=20pt,bblly=221pt,bburx=570pt,bbury=608pt,width=8.9cm]
{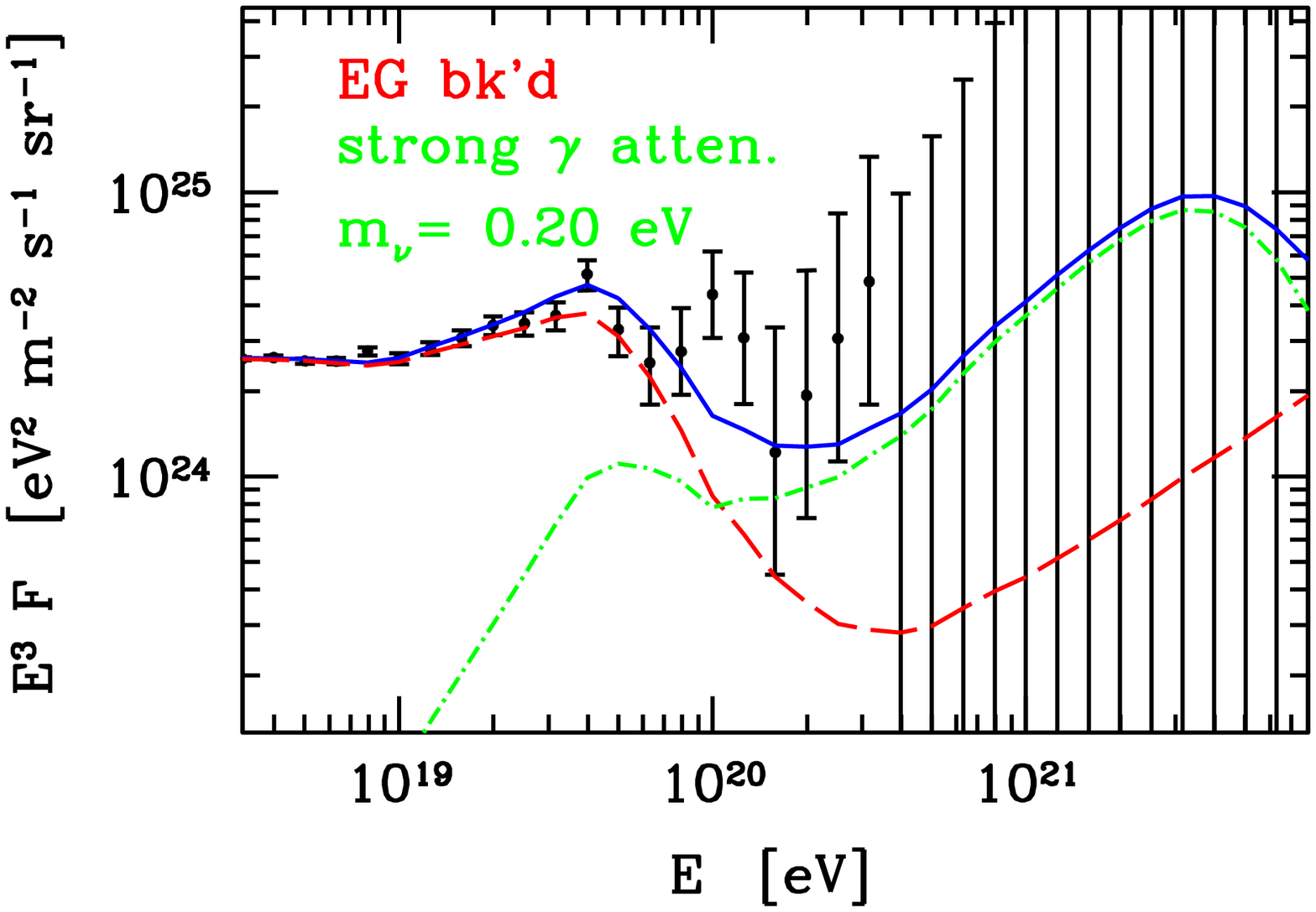}\\
\includegraphics[bbllx=20pt,bblly=221pt,bburx=570pt,bbury=608pt,width=7.7cm]
{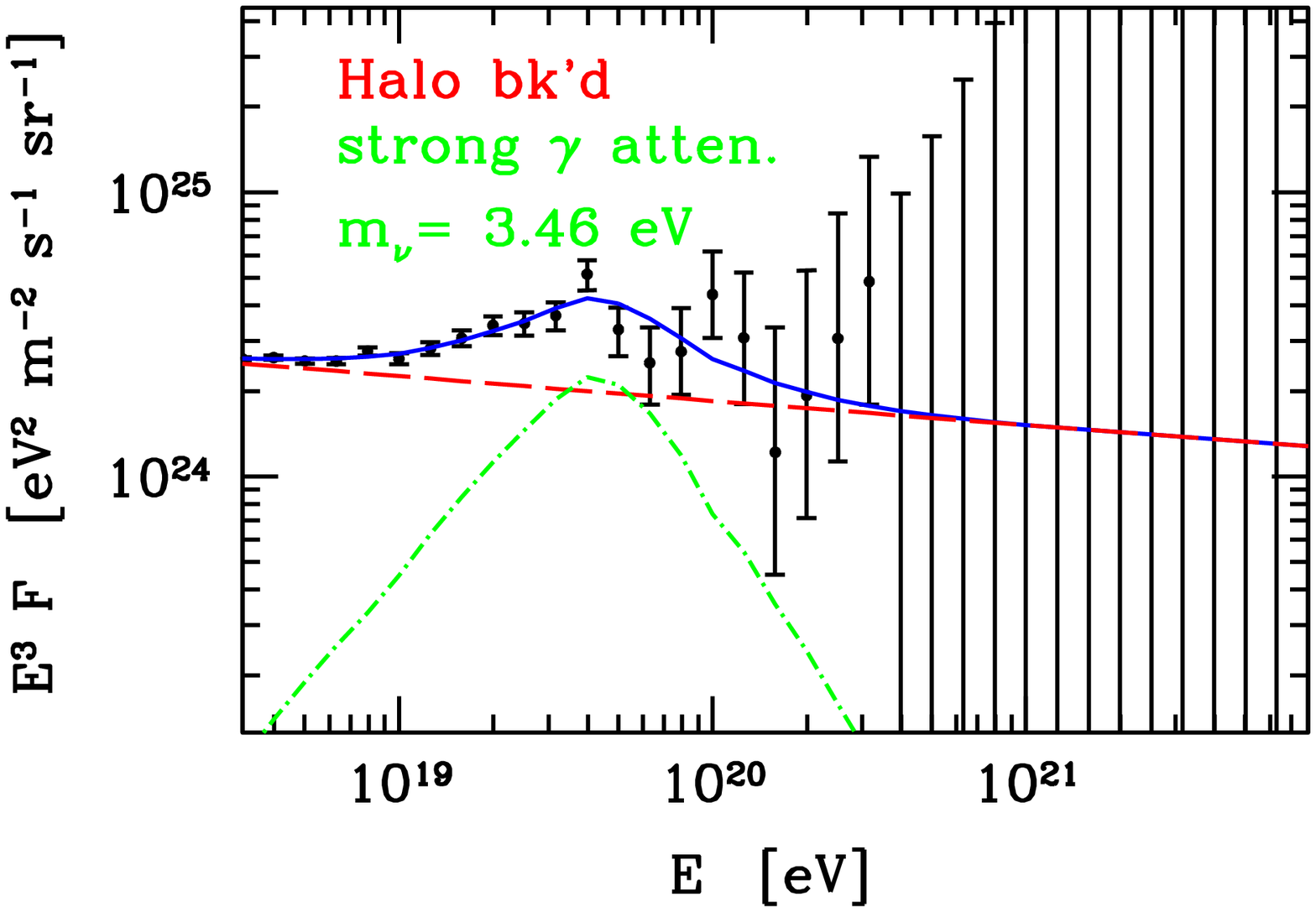}
\includegraphics[bbllx=20pt,bblly=221pt,bburx=570pt,bbury=608pt,width=7.7cm]
{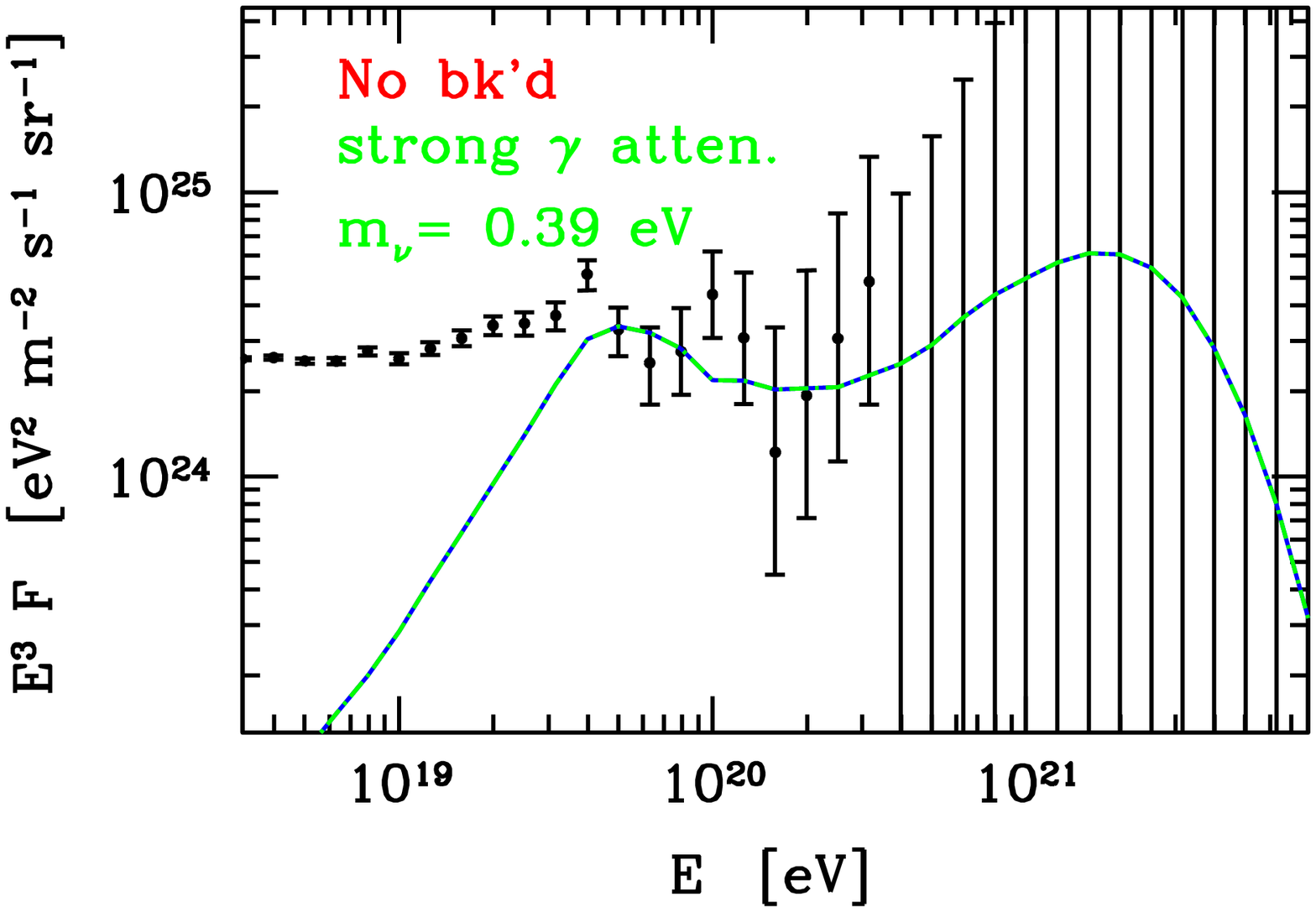}
\caption[...]{\label{fit_nu}
The available UHECR data with their error bars
and the best fits from Z-bursts, for 
a strong UHE$\gamma$ attenuation such that
the Z-burst photons can be neglected ($\alpha =0, h=0.71, \Omega_M= 
0.3,\Omega_\Lambda =0.7,z_{\rm max}=2$).
{\em Top:} 
The case of an ``extragalactic'' UHECR background.
The first bump at $4\cdot 10^{19}$~eV represents protons produced at
high energies and accumulated just above the GZK cutoff due to their energy
losses. The bump at $2\cdot 10^{21}$~eV is a remnant of the Z-burst
energy. The long-dashed line shows the contribution of the power-law-like
spectrum with the GZK effect included. 
{\em Bottom left :} 
Best fit for the case of a halo background (solid line). The bump 
around $4\cdot 10^{19}$~eV is mainly due to the Z-burst protons (dash-dotted), 
whereas 
the 
almost horizontal contribution (long-dashed) is the first, power-law-like 
term of Eq.~(\ref{flux}). 
{\em Bottom right:} 
The case of no UHECR background above $\log (E_{\rm min}/\mbox{eV})= 19.7$. 
}
\end{center}
\end{figure}

\begin{table}
\caption{\label{fit_noph}Results of fits, for 
a strong UHE$\gamma$ attenuation ($h=0.71,\Omega_M= 0.3,\Omega_\Lambda 
=0.7,z_{\rm max}=2$). 
{\em Top:} Assuming a halo UHECR background according to 
Eq.~(\ref{pow-law-halo}).
{\em Middle:} Assuming an extragalactic UHECR background according to 
Eq.~(\ref{pow-law-eg}).
{\em Bottom:} Assuming no UHECR background above $E_{\rm min}$, for different 
values of the 
lower end $E_{\rm min}$ of the fit ($\alpha = 0$).\\}

\begin{tabular*}{\textwidth}{@{}c*{15}{@{\extracolsep{0pt plus12pt}}c}}
\hline
\multicolumn{6}{c}{EG UHECR background + strong UHE$\gamma$ attenuation}\\ 
$\alpha$ & $m_\nu$ [eV] & $\chi^2_{\rm min}$ & $A$ & $B$ & $\beta$\\ \hline
$-3$&   $0.20^{+0.20(0.63)}_{-0.11(0.18)}$&     $25.82$&        $5.00\cdot 
10^{31}$&    $150$&  $2.465$\\
$0$&    $0.20^{+0.19(0.61)}_{-0.12(0.18)}$&     $26.41$&        $5.98\cdot 
10^{31}$&    $144$&  $2.466$\\
$3$&    $0.20^{+0.19(0.59)}_{-0.11(0.17)}$&     $26.89$&        $7.23\cdot 
10^{31}$&    $142$&  $2.467$\\
\hline
\multicolumn{6}{c}{Halo UHECR background + strong UHE$\gamma$ attenuation}\\ 
$\alpha$ & $m_\nu$ [eV] & $\chi^2_{\rm min}$ & $A$ & $B$ & $\beta$\\ \hline
$-3$&   $4.46^{+2.22(4.80)}_{-1.64(2.88)}$&     $15.90$&        $1.46\cdot 
10^{43}$&    $1049$& $3.110$\\
$0$&    $3.46^{+1.73(4.03)}_{-1.34(2.32)}$&     $15.64$&        $1.62\cdot 
10^{43}$&    $770$&  $3.111$\\
$3$&    $2.51^{+1.45(3.30)}_{-1.05(1.80)}$&     $15.53$&        $1.65\cdot 
10^{43}$&    $551$&  $3.111$\\
\hline
\multicolumn{6}{c}{No UHECR background + strong UHE$\gamma$ attenuation}\\ 
$\log (E_{\rm min}/\mbox{eV})$ & $m_\nu$ [eV] & $\chi^2_{\rm min}$ & $A$ & $B$ 
& $\beta$\\ \hline
$19.4$& $2.28^{+0.64(1.46)}_{-0.58(1.06)}$&     $21.81$&        $-$&    $1251$& 
$-$\\
$19.5$& $1.31^{+0.63(1.44)}_{-0.53(0.80)}$&     $16.01$&        $-$&    $846$&  
$-$\\
$19.6$& $0.85^{+0.67(1.62)}_{-0.31(0.55)}$&     $14.80$&        $-$&    $670$&  
$-$\\
$19.7$& $0.40^{+0.32(0.87)}_{-0.16(0.27)}$&     $8.03$& $-$&    $445$&  $-$\\
$19.8$& $0.42^{+0.41(1.25)}_{-0.18(0.29)}$&     $7.99$& $-$&    $460$&  $-$\\
$19.9$& $0.76^{+1.06(2.50)}_{-0.39(0.58)}$&     $5.52$& $-$&    $733$&  $-$\\
$20.0$& $1.77^{+1.49(3.47)}_{-1.01(1.47)}$&     $2.68$& $-$&    $2021$& $-$\\
\hline
\end{tabular*}
\end{table}

As already mentioned, presently there is no evidence that the observed highest 
energy cosmic
rays are photons. Let us start therefore with the assumption (cf. 
Ref.~\cite{Fodor:2001qy}) 
that the ultrahigh energy photons from 
Z-bursts can be neglected in the fit in comparison to the protons. This is 
certainly true
for a sufficiently large universal radio background, 
and/or for a sufficiently strong 
extragalactic 
magnetic field ${\mathcal O}(10^{-9})$~G. 
We shall refer to this scenario in the following as ``strong'' UHE$\gamma$ 
attenuation.  
Our best fits to the observed data for this scenario are given by  
Table~\ref{fit_noph}. We can see that the required neutrino masses
are rather insensitive to the evolution parameter $\alpha$. 
The fits can be seen in 
Fig.~\ref{fit_nu}, for $\alpha=0$. 
We found a neutrino mass of
\massnophhalo
for the ``halo background'' scenario, \massnopheg   
if the background is extragalactic, and 
\massnophno
if there are are no background protons above 
$10^{19.7}$~eV. The first numbers are
the 1\,$\sigma$, the numbers in the brackets are the
2\,$\sigma$ errors. This gives an absolute lower bound on the mass of the 
heaviest neutrino of 
\lowbdnopheg
at the 95\% C.L., which is comparable to the one obtained from the atmospheric
mass splitting in a three flavour scenario, Eq.~(\ref{lim_low_atm}).

\begin{figure}
\begin{center}
\includegraphics[bbllx=20pt,bblly=221pt,bburx=570pt,bbury=608pt,width=8.9cm]
{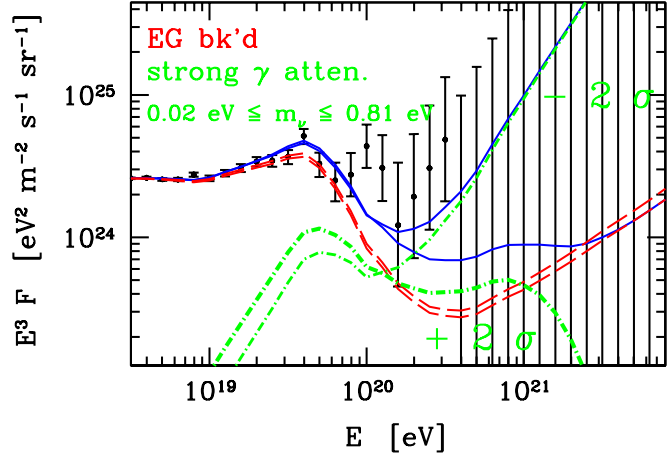}
\caption{ \label{fit_errors}
The available UHECR data with their error bars
and the $\chi^2_{\rm min}\pm 4$ fits (solid lines) from Z-bursts
for  an extragalactic 
UHECR background (long-dashed lines), and 
a strong UHE$\gamma$ attenuation  such that
the Z-burst photons can be neglected 
and only the Z-burst protons (dash-dotted lines) have to be taken into account 
in the fit 
($\alpha =0, h=0.71, \Omega_M= 0.3,\Omega_\Lambda =0.7,z_{\rm max}=2$).}
\end{center}
\end{figure}

The surprisingly small uncertainties are based on the 
$\chi^2$ analysis described previously.  
The inclusion of the already mentioned 30\,\% uncertainties in the observed 
energies
by a Monte Carlo analysis increases the error bars by about 10\,\%. 
Note, that the relative errors in the extragalactic and in the no 
background cases are of the same order. This clearly
shows that the smallness of these
errors does not originate from the 
low energy part of the background component.

The fits are rather good:
for 21 non-vanishing bins and 4 fitted parameters they can be as low as 
\chiminnophhalo
and 
\chiminnopheg 
in the halo and the extralactic background case, respectively,  
whereas in the no background case, for $E_{\rm min}=10^{19.7}$~eV, we have 
9 bins with 2 fitted parameters and a 
\chiminnophno
see Table~\ref{fit_noph}.  
In the latter case, however, 
a remarkable
dependence of the fitted mass on the value of $E_{\rm min}$ is observed. 
The $\pm\, 2\ \sigma$ fits are shown in
Fig.~\ref{fit_errors} for the extragalactic case. 
According to the expectations the spread is small in the region where 
there are data, whereas it can be  quite large in the presently unexplored 
ultrahigh energy regime.

\begin{figure}
\begin{center}
\includegraphics[bbllx=20pt,bblly=221pt,bburx=570pt,bbury=608pt,width=8.9cm]
{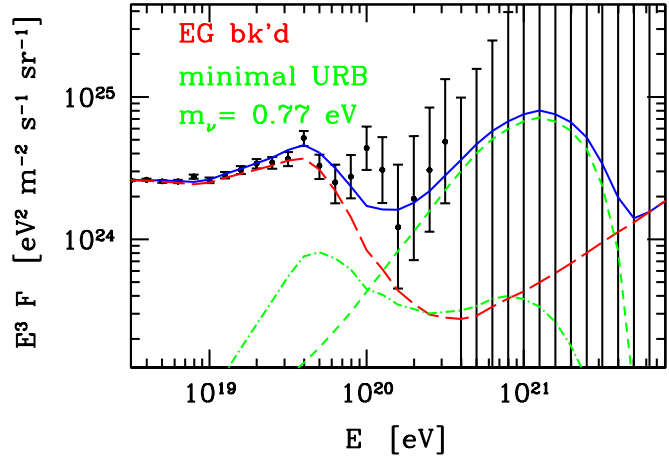}
\caption[...]{\label{fit_nu_minurb}
The available UHECR data with their error bars
and the best fit from Z-bursts, 
for an energy attenuation corresponding to ``minimal'' URB  
and a vanishing EGMF in the case of extragalactic background. 
($\alpha=0,h=0.71,\Omega_M= 0.3,\Omega_\Lambda =0.7,z_{\rm max}=2$).
The best fit (solid line) is the sum of the contributions of 
the 
background protons (long-dashed), the Z-burst protons (dash-dotted)
and the Z-burst photons (short-dashed).}
\end{center}
\end{figure}

As the other extreme, let us consider now the case of ``minimal'' URB 
on the level of the one in~\cite{Clark:1970} and
a vanishing EGMF. Then the photons will give a non-negligible contribution
in the ultrahigh energy region.
The corresponding fit results for $\alpha=0$ 
are given by Table~\ref{fit_minurb} while
the best fit for the extragalactic background case is shown by
Fig.~\ref{fit_nu_minurb}.
The values of the neutrino mass found in both the halo 
background scenario, 
with 
\massleehalo
 as well as in the extragalactic background scenario, 
with 
\massleeeg
 are compatible 
with the corresponding values found in the case of strong UHE$\gamma$ 
attenuation. 

\begin{table}
\caption{\label{fit_minurb}Results of fits, for 
a photon attenuation corresponding to ``minimal'' URB and 
vanishing EGMF. 
($\alpha=0,h=0.71,\Omega_M= 0.3,\Omega_\Lambda =0.7,z_{\rm max}=2$).\\}  
\begin{tabular*}{\textwidth}{@{}c*{15}{@{\extracolsep{0pt plus12pt}}c}}
\hline
${\rm background}$ & $m_\nu$ [eV] & $\chi^2_{\rm min}$ & $A$ & $B$ & $\beta$\\ \hline
EG&    $0.77^{+0.48(1.36)}_{-0.30(0.51)}$&     $24.52$&        $6.44\cdot 
10^{31}$&    $120$&  $2.467$\\
Halo&    $3.71^{+1.40(3.27)}_{-1.12(1.96)}$&     $15.36$&        $1.05\cdot 
10^{44}$&    $827$&  $3.166$\\
\hline
\end{tabular*}
\end{table}

\begin{figure}\begin{center}
\includegraphics[bbllx=20pt,bblly=221pt,bburx=570pt,bbury=608pt,width=8.65cm]
{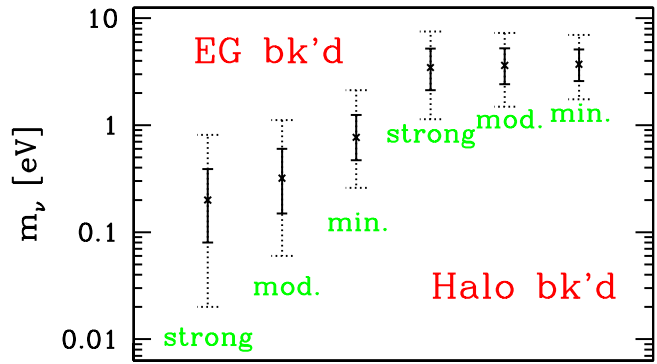}
\caption[...]{\label{mass_res}
Summary of the masses of the heaviest neutrino required in the Z-burst 
scenario, with
their 1\,$\sigma$ (solid) and 2\,$\sigma$ (dotted) error bars, for the case of 
an extragalactic and a halo  
background of ordinary cosmic rays and   
for various assumptions about the diffuse extragalactic photon background in 
the radio band 
($\alpha =0, h=0.71, \Omega_M= 0.3,\Omega_\Lambda =0.7,z_{\rm max}=2$).
From left: strong $\gamma$ attenuation, moderate and minimal URB.
}
\end{center}\end{figure}

We also performed the fits for other values of the
cosmological parameters and relic neutrino overdensities.
We found that for a wide range 
(e.g. $\alpha =-3\div 3$, $h=0.61\div 0.9$, $z_{\rm 
max}=2\div 5$) the results remain well within the errorbars.  
The only dependence (still within the errorbars) 
is caused by the choice of the background
and the type of photon propagation 
(i. e. the strength of the URB and/or EGMF).
Fig.~\ref{mass_res} contains a summary of the neutrino masses for
the extragalactic and halo scenarios and three different photon propagations
(the already mentioned strong UHE$\gamma$ attenuation and the ``minimal'' 
URB and a ``moderate`` URB case in between the two extremes).
The overall mass range for the heaviest neutrino in the case of halo 
background is
\massrangehaloonesig
at the 68\,\% C.L., if we take into account the 
variations between the
minimal and moderate URB cases and the strong UHE$\gamma$ attenuation case.
For the extragalactic background scenario, the required mass range is
\massrangeegonesig.

\begin{figure}\begin{center}
\includegraphics[bbllx=20pt,bblly=221pt,bburx=570pt,bbury=608pt,width=8.65cm]
{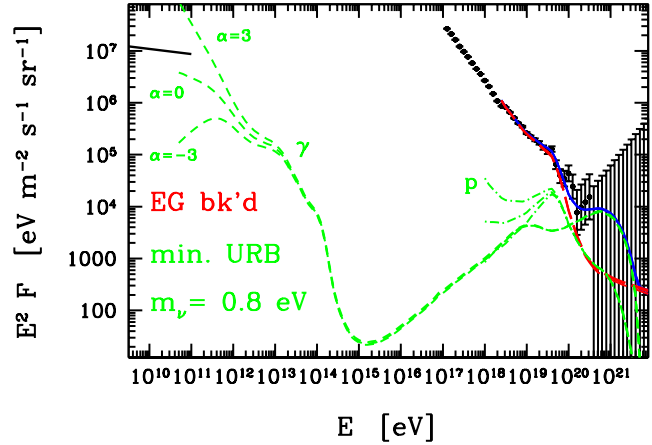}
\caption[...]{\label{fit_egret}
The available UHECR data with their error bars
and the best fit from Z-bursts, 
for various cosmological evolution parameters $\alpha$ and 
an energy attenuation of photons as in Fig.~\ref{ph_mfp} (bottom) exploiting  
a ``minimal'' URB  
($h=0.71, \Omega_M= 0.3,\Omega_\Lambda =0.7,z_{\rm max}=2$).  
Also shown is the diffuse $\gamma$ background in the energy range 
between 30 MeV and 100 GeV as measured by EGRET (solid). 
}
\end{center}
\end{figure}

Let us consider in more detail the $\gamma$ ray spectra from Z-bursts, 
notably
in the $\sim 100$~GeV region. As illustrated in Fig.~\ref{fit_egret}, 
the EGRET measurements of the diffuse $\gamma$ background in the energy range 
between 30~MeV and 100~GeV~\cite{Sreekumar:1998} gives a non-trivial constraints
on the evolution parameter $\alpha$. 
Whereas the high energy spectrum, and thus the neutrino mass,
is independent of $\alpha$, at low energies only
$\alpha\lwig 0$ seems to 
be compatible with the EGRET measurements, quite 
independently of different 
assumptions about the URB.      
These numerical findings are in fairly good agreement with other recent 
simulations~\cite{Kalashev:2001sh}. These photon fluxes do not contain any
contribution from direct photons emitted by the UHE$\nu$ sources. As they are
already close to the EGRET limit, one needs special sources that do not give 
contribution to the EGRET region.

\begin{figure}\begin{center}
\includegraphics[bbllx=20pt,bblly=221pt,bburx=570pt,bbury=608pt,width=8.65cm]
{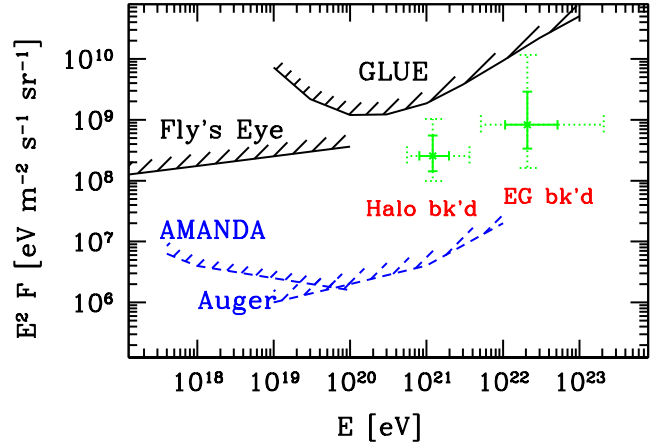}
\caption[...]{\label{eflux}
Neutrino fluxes, 
$F = \frac{1}{3} \sum_{i=1}^3 ( F_{\nu_i}+F_{\bar\nu_i})$, required 
by the Z-burst hypothesis for the case of a halo and an extragalactic 
background of 
ordinary cosmic rays, respectively ($\alpha =0, h=0.71, \Omega_M= 
0.3,\Omega_\Lambda =0.7,z_{\rm max}=2$).
Shown are the necessary fluxes obtained from the fit results of 
Table~\ref{fit_noph} 
for the case of a strong UHE$\gamma$ attenuation.
The horizontal errors indicate the 1\,$\sigma$ (solid) and 2\,$\sigma$ (dotted) 
uncertainties of the
mass determination and the vertical errors include also the uncertainty
of the Hubble expansion rate.
Also shown are upper limits from Fly's Eye on 
$F_{\nu_e}+F_{\bar\nu_e}$ and GLUE on $\sum_{\alpha = e,\mu} ( 
F_{\nu_\alpha}+F_{\bar\nu_\alpha})$, 
as well as projected sen\-si\-tivi\-ties of 
AMAN\-DA on $F_{\nu_\mu}+F_{\bar\nu_\mu}$ and 
Auger on $F_{\nu_e}+F_{\bar\nu_e}$. The 
sensitiviy of RICE is comparable to 
the one of Auger.}
\end{center}
\end{figure}

The necessary UHEC$\nu$ flux at the resonant energy $E_\nu^{\rm res}$ is 
shown by Fig.~\ref{eflux} together with the existing experimental limits and
the projected sensitivities of present and future observational projets.
They appear to be well below present upper limits and are within the expected 
sensitivity of 
AMANDA~\cite{Barwick00,Hundertmark:2001},  RICE~\cite{Seckel:2001}, and 
Auger~\cite{Capelle98}.  
The fluxes are consistent with the ones found in 
Refs.~\cite{Kalashev:2001sh,Gelmini:2002xy}. 

The astrophysical sources for the required UHE neutrinos should be 
distributed with
$\alpha \lwig 0$, accelerate protons up to energies $\gwig\, 10^{23}$~eV, be opaque 
to primary
nucleons and emit secondary photons only in the sub-MeV region. 
It is an interesting question whether such challenging conditions can be 
realized in BL Lac objects, a 
class of active galactic nuclei for which some evidence of zero or negative 
cosmological evolution has been 
found (see Ref.~\cite{Caccianiga:2001} and references therein) and which were 
recently discussed as 
possible sources of the highest energy cosmic 
rays~\cite{Tinyakov:2001nr,Gorbunov:2002nb,Neronov:2002xv}.

It should be stressed that, besides the neutrino mass, the UHEC$\nu$ flux at 
the resonance 
energy is one of the most robust predictions of the Z-burst scenario which can 
be verified
or falsified in the near future. 

\section{\label{sect:discussion}Conclusions}

We have presented a  comparison of the predicted spectra from 
Z-bu\-rsts -- resulting from the resonant annihilation of ultrahigh energy 
cosmic neutrinos
with relic neutrinos --  
with the observed ultrahigh energy cosmic ray spectrum.
The mass of the heaviest neutrino should be in the
range \massrangehaloonesig 
on the 1~$\sigma$ level, if 
the background of ordinary cosmic rays above $10^{18.5}$~eV consists of protons 
and 
originates from a region within a distance of about 50~Mpc. 
In the phenomenologically most plausible case that the ordinary cosmic rays 
above $10^{18.5}$ are 
protons of extragalactic origin the required neutrino mass range is 
\massrangeegonesig\ at the 68\,\% C.L.. 
In the case with no background of ordinary cosmic 
rays above, say, $10^{19.7}$~eV we found \massrangenoonesig.
The above neutrino mass ranges include variations in presently 
unknown quantities,  like the amount of neutrino clustering, the universal 
radio background, and the
extragalactic magnetic field, within their anticipated uncertainties. 
There is experimental indication that the highest energy events are mostly
not photons. 
In this special case, when all ultrahigh energy photons are suppressed by a
strong enough URB and/or EGMF, the mass range narrows down (
with an extragalactic background) to
\massrangeegstrongonesig, with a best 
fit value of \massnophegbestfit.  

It is remarkable, that the mass ranges required for the Z-burst scenario 
coincide nearly perfectly
with the present knowledge about the mass of the heaviest neutrino from 
oscillations and tritium
$\beta$ decay , 
$0.04\ {\rm eV}\,\lwig\, m_{\nu_3}\,\lwig\,2.5$~eV, 
in a three flavour, or 
$0.4\ {\rm eV}\,\lwig\, m_{\nu_4}\,\lwig\,3.8$~eV, in a four flavour scenario.  
   
It is interesting to 
observe that the recently
reported evidence for neutrinoless double beta 
decay~\cite{Klapdor-Kleingrothaus:2002ke}, with   
$0.11\ {\rm eV}\leq\langle m_\nu\rangle\leq 0.56$~eV,
if true, is compatible with our ``favourite'' 
extragalactic/strong URB scenario. 

We also determined the necessary ultrahigh energy neutrino 
flux at the 
resonance energy. 
It was found   
to be consistent with present upper limits and detectable in the near future by 
the already 
operating neutrino telescopes AMANDA and RICE, and by the Pierre Auger 
Observatory.
A search at these facilities is the most promising test of
the Z-burst hypothesis.   

The required neutrino fluxes are enormous. 
If such tremendous fluxes of ultrahigh energy neutrinos are indeed found, one 
has to deal with
the challenge to explain their origin. 
It is fair to say, that at the moment no 
convincing astrophysical
sources are known which meet the requirements for the Z-burst hypothesis, i.e. 
which  
have no or a negative cosmological evolution, 
accelerate protons at least up to $10^{23}$~eV, are opaque to primary
nucleons and emit secondary photons only in the sub-MeV region. 

\section*{Acknowledgments}
We thank the OPAL collaboration for their unpublished
results on hadronic Z decays.
This work was partially supported by Hungarian Science Foundation
grants No.
OTKA-T37615/\-T34980/\-T29803/\-M37071/\-OMFB1548/\-OMMU-708.


\end{document}